\RequirePackage{fix-cm}
\documentclass[smallextended]{svjour3}       
\smartqed  
\usepackage{graphicx}
\usepackage{amsmath,epsfig,amsfonts,dsfont,color}

\newcommand{\RR}{{\rm I\!R}}

\def\x{{\mathbf x}}
\def\y{{\mathbf y}}
\def\z{{\mathbf z}}

\def\r{{\mathbf r}}

\begin{document}

\title{Bayesian Conditional Monte Carlo Algorithms for Sequential Single and Multi-Object filtering 
}



\author{Yohan Petetin, Fran\c{c}ois Desbouvries}
\institute{Telecom and Mines Institute / Telecom SudParis / CITI Department and CNRS UMR 5157, 
              9 rue Charles Fourier, 91011 Evry, France.
              Email: {\tt \{yohan.petetin,francois.desbouvries\}@telecom-sudparis.eu} }

\date{Received: date / Accepted: date}

\maketitle

\begin{abstract}
Bayesian filtering aims at tracking sequentially
a hidden process from an observed one.
In particular,
sequential Monte Carlo (SMC) techniques
propagate in time weighted trajectories
which represent the posterior probability density function (pdf)
of the hidden process given the available observations.
On the other hand,
Conditional Monte Carlo (CMC)
is a variance reduction technique
which replaces the estimator of a moment of interest
by its conditional expectation given another variable.
In this paper we show that up to some adaptations,
one can 
make use of the time recursive nature
of SMC algorithms
in order to propose natural temporal CMC
estimators of some point estimates of the hidden process,
which outperform the associated crude Monte Carlo (MC) estimator
whatever the number of samples.
We next 
show that
our Bayesian CMC estimators can be computed exactly,
or approximated efficiently,
in some hidden Markov chain (HMC) models;
in some jump Markov state-space systems (JMSS);
as well as in multitarget filtering.
Finally our algorithms are validated via simulations. 
\end{abstract}

\section{Introduction}
\label{sec:intro}

\subsection{SMC algorithms for single- or multi-object Bayesian filtering}

In single object Bayesian filtering we consider two random processes 
$\{X_n\}_{n\geq 0}$ and
$\{Y_n\}_{n\geq 0}$
with given joint probability law.
$Y_i$ is observed, i.e.
we have at our disposal realizations 
${\bf y}_{0:n} = \{ y_i  \}_{i=0}^n$ of
${\bf Y}_{0:n} = \{ Y_i  \}_{i=0}^n$
(as far as notations are concerned,
upper case letters denote random variables (r.v.),
lower case ones their realizations,
and bold letters vectors;
$p(x)$, say, denotes the pdf
of r.v. $X$ and 
$p(x|y)$, say, the conditional pdf of $X$ given $Y=y$;
if $i \leq j$ $p_{i:j|n}$ 
is a shorthand notation for 
$p({\bf x}_{i:j}|{\bf y}_{0:n})$;
if $x^{i}$ are samples from $p(x)$
then the set 
$\{ x^i \}_{i=1}^N$ can also be denoted 
${\bf x}^{1:N}$;
subscripts are reserved for times indices and
superscripts for realizations).
Process $\{ X_n \}$ is hidden,
and our aim is to compute,
for each time instant $n$,
some moment of interest 
\begin{equation}
\label{Theta_n-def-generale}
\Theta_n = \int f({\bf x}_{0:n}) p({\bf x}_{0:n}|{\bf y}_{0:n}) {\rm d}{\bf x}_{0:n}
\end{equation}
of the a posteriori 
pdf
$p({\bf x}_{0:n}|{\bf y}_{0:n})$ of 
${\bf X}_{0:n}$ given
${\bf y}_{0:n}$. 
Unfortunately, in most models 
\eqref{Theta_n-def-generale}
cannot be computed exactly. 
Suboptimal solutions for computing $\Theta_n$
include 
SMC techniques
\cite{livredoucet}
\cite{arulampalamshort},
which propagate over time 
weighted trajectories
$\{{\bf x}_{0:n}^i,w_n^i\}_{i=1}^N$
with 
$\sum_{i=1}^N w_n^i = 1$.
In other words,
$\widehat{p}_{0:n|n} = \sum_{i=1}^N w_n^i \delta_{{\bf x}_{0:n}^i}$,
in which $\delta$ is the Dirac mass,
is a discrete (and random) approximation
of $p({\bf x}_{0:n}|{\bf y}_{0:n})$.

On the other hand, 
{\sl multi}-object filtering (see e.g. \cite{MAHLER_ARTICLE2003})
essentially reduces to computing 
$\Theta_n = \int f({\bf x}_{n}) v_n({\bf x}_{n}) {\rm d}{\bf x}_{n}$
in which
$v_n({\bf x}_{n})$ is now 
the so-called Probability Hypothesis Density (PHD), 
i.e. the a posteriori spatial density
of the expected number of targets,
given all measurements
(be they due to detected targets or to false alarms).
Again, SMC techniques propagate an approximation
of $v_n$ with a set of weighted samples
$\{{\bf x}_{n}^i,w_n^i\}_{i=1}^N$; 
here $\sum_{i=1}^N w_n^i$, which in general is different from $1$,
is an estimator of the number of targets.

Now, SMC algorithms, be they for single- or multi-object Bayesian filtering,
usually focus on how to propagate 
approximations $\widehat{p}_{0:n|n}$ (or $\widehat{v}_{n}$)
of ${p}_{0:n|n}$ (or $v_n$);
once $\widehat{p}_{0:n|n}$ or $\widehat{v}_{n}$ has been computed,
$\Theta_n$ is finally estimated either as
$\widehat{\Theta}_n = \sum_{i=1}^N w_n^i f({\bf x}_{0:n}^i)$
or $\sum_{i=1}^N w_n^i f({\bf x}_{n}^i)$.
By contrast, in this paper we directly focus 
on $\widehat{\Theta}_n$ itself,
and see under which conditions one can improve this point estimator
at a reasonable computational cost.

\subsection{Variance reduction via conditioning: Rao-Blackwellized particle filters (RB-PF)}
\label{intro-variance-reduction}
This problem leads us to variance reduction techniques which form
an important part of computer simulation
(see e.g. \cite{asmussen-glynn}).
Among them,
methods based on conditioning variables
rely on the following well known result.
Let $X_1$ and $X_2$ be two r.v.
and $f$ some function.
Then
\begin{eqnarray}
\label{egalite-esperances-conditionnel}
{\rm E}({\rm E}(f(X_2)|X_1)) & = & {\rm E}(f(X_2)), 
\\
\label{inegalite-variances-RB}
{\rm var}({\rm E}(f(X_2)|X_1))
& = &
{\rm var}(f(X_2))
-
{\rm E}({\rm var}(f(X_2)|X_1)).
\end{eqnarray}
So if the aim is to compute $\Theta={\rm E}(f(X_2))$
and we have at our disposal 
$\{X_1^i\}_{i=1}^N \stackrel{\rm i.i.d.}{\sim} p(x_1)$, 
$\{X_2^i\}_{i=1}^N \stackrel{\rm i.i.d.}{\sim} p(x_2)$
then the so-called CMC estimator
$\widetilde{\Theta}  = 
\frac{1}{N}
\sum_{i=1}^N
{\rm E}(f(X_2)|X_1^i)$
has lower variance than the
corresponding crude MC one
$
\widehat{\Theta}  = 
\frac{1}{N}
\sum_{i=1}^N
f(X_2^i).
$ 
Of course, the interest 
of $\widetilde{\Theta}$ 
vs. $\widehat{\Theta}$ 
depends on the choice of $X_1$:
ideally, one should easily sample from $p(x_1)$;
the variance reduction in \eqref{inegalite-variances-RB}
should be as large as possible;
but in the meantime function $g(x_1) = {\rm E}(f(X_2)|x_1)$
should remain computable at a reasonable computational cost.

Variance reduction techniques based on 
CMC methods have been adapted
to Bayesian filtering;
in this context, 
these methods are either known as marginalized or RB-PF
\cite{Chen-Mixture-Kalman}
\cite{doucet-sequentialMC} 
\cite{doucet-jump-Mkv}
\cite{Shon-MPF}.
The rationale is as follow.
Let now $\Theta_n$ in \eqref{Theta_n-def-generale}
be rewritten as
$\Theta={\rm E}(f(X_1,X_2))$.
It is usually not possible to sample from $p(x_1,x_2)$,
and often $p(x_1,x_2) \propto p'(x_1,x_2)$ 
is only known up to a constant, 
whence the use of 
Bayesian (or normalized) importance sampling (IS) techniques 
\cite{Gewecke}. 
So let
%
%
\begin{align}
\label{theta-smc-1}
\widehat{\Theta}({\bf x}_1^{1:N},{\bf x}_2^{1:N}) &= 
\sum_{i=1}^N
w_2^i({\bf x}_1^{1:N},{\bf x}_2^{1:N})
f(x_1^i,x_2^i) 
\text{ with } 
(x_1^i,x_2^i) 
\sim
q_2 \text{,}   
\\
\label{theta-smc-2}
\widehat{\Theta}^{\rm RB}({\bf x}_1^{1:N})&= 
\sum_{i=1}^N
w_1^i({\bf x}_1^{1:N})
{\rm E}(f(x_1^i,X_2)|x_1^i) 
\text{ with } 
x_1^i 
\sim
q_1,
\end{align}
with 
$\sum_{i=1}^N w_1^i =$
$\sum_{i=1}^N w_2^i = 1.$
Estimator $\widehat{\Theta}^{\rm RB}$
depends on samples $\{x_1^i\}_{i=1}^N$ only 
and is known as the RB estimator of $\Theta$.
However
$\widehat{\Theta}^{\rm RB}$
is known to outperform 
$\widehat{\Theta}$
only under specific assumptions 
on $q_1$, $q_2$,
${\bf w}_1^{1:N}$ and
${\bf w}_2^{1:N}$.
In particular, 
if
$w_1^i \propto w_1^{u,i} = p'(x_1^i)/q_1(x_1^i)$,
$w_2^i \propto w_2^{u,i} = p'(x_1^i,x_2^i)/q_2(x_1^i,x_2^i)$ and
$q_1(x_1) = \int q_2(x_1,x_2){\rm d}x_2$,
then the variance of
$w_1^{u,i}$ can only lower than that of $w_2^{u,i}$
\cite{doucet-sequentialMC}.
If moreover 
$(x_1^i,x_2^i)$ are independent,
an asymptotic analysis 
based on 
\eqref{egalite-esperances-conditionnel} and
\eqref{inegalite-variances-RB}
proves that 
$\widehat{\Theta}^{\rm RB}$ 
indeed outperforms $\widehat{\Theta}$
\cite{doucet-jump-Mkv}.
However, independence never holds in the presence of resampling;
in the general case, 
the comparison of 
both estimators depends on the choice of the 
importance distributions 
$q_1$ and $q_2$,
and can be proved (asympotically) 
only under specific sufficient conditions
\cite{chopin}
\cite{lindstein-rao}.

RB-PF have been applied in the specific
case where the state vectors $\x_{0:n}$ can be partitioned 
into a ``linear'' component $\x_2=\x_{0:n}^l$ 
and a ``non-linear'' one $\x_1=\x_{0:n}^{nl}$.
Models in which computing 
$\widehat{\Theta}^{\rm RB}$ is possible 
include linear and Gaussian JMSS 
\cite{doucet-jump-Mkv} \cite{Chen-Mixture-Kalman}
or partially linear and Gaussian HMC     
\cite{Shon-MPF}.
In other models, it may be possible to 
approximate $\widehat{\Theta}^{\rm RB}$ by using
numerical approximations of $w_1(x)$ and of
${\rm E}(f(X_1,X_2)|x_1)$. 
However, due to 
the spatial structure of the decomposition of $\x_{0:n}$,
approximating $\Theta_n$ in \eqref{Theta_n-def-generale}
involves propagating numerical approximations over time.

\subsection{Bayesian CMC estimators}
\label{bayesian-cmc-estimators}

\subsubsection{Spatial vs. temporal RB-PF}

In this paper we propose another class of RB-PF;
the main difference is that our partitioning 
$(X_1,X_2)$ 
of $\x_{0:n}$ is now temporal rather than spatial.
The question arises naturally in the Bayesian filtering context:
at time $n$ we usually build $\Theta_n$ from
$\widehat{p}_{0:n|n}$,
but indeed 
$\widehat{p}_{0:n-1|n-1}$ was also available for free
since, by nature, {\sl sequential} MC algorithms construct 
$\widehat{p}_{0:n|n}$ {\sl from}
$\widehat{p}_{0:n-1|n-1}$.
Now, comparing with spatially partitioned RB-PF,
a temporal partition of $\x_{0:n}$ 
has a number of statistical and computational
structural consequences,
as we now see. 
So let again 
\begin{eqnarray}
\label{def-de-Theta-jointe}
\Theta & = & 
\int f(x_1,x_2) p(x_1,x_2) {\rm d}x_1{\rm d}x_2 \\
& = &
\label{def-de-Theta-conditionnel}
\int \left[ \int f(x_1,x_2) p(x_2|x_1) {\rm d}x_2 \right] p(x_1) {\rm d}x_1.
\end{eqnarray}
Let us start from the following approximation of $p(x_1)$:
\begin{equation}
\label{approx-p1}
p(x_1) \approx
\widehat{p}(x_1) = \sum_{i=1}^N w^i({\bf x}_1^{1:N}) \delta_{x_1^i}.
\end{equation}
For $1 \leq i \leq N$ let next $x_2^i \sim p(x_2|x_1^i)$.
This yields the following approximation of $p(x_1,x_2)$:
\begin{equation}
\label{approx-p1p2}
\widehat{p}(x_1,x_2)  = \sum_{i=1}^N w^i({\bf x}_1^{1:N}) \delta_{(x_1^i,x_2^i)};
\end{equation}
note that each weight $w^i$ may depend on 
$\{x_1^i\}_{i=1}^N$, but not on 
$\{x_2^i\}_{i=1}^N$.
The reason why is that 
we now use a temporal partition,
and not
a spatial one: 
in the spatial subdivision case,
$p(x_2|x_1)$ would reduce to
$p(\x_{0:n}^l|\x_{0:n}^{nl},\y_{0:n})$,
which means that we would need to sample at each time step the whole
set $\{\x_{0:n}^{l,i}\}_{i=1}^N$,
instead of simply extending the trajectories.

Finally 
we have two options: 
computing 
the full expectation in  
\eqref{def-de-Theta-jointe} by using 
\eqref{approx-p1p2},
or only the outer one 
in \eqref{def-de-Theta-conditionnel} by using 
\eqref{approx-p1}. So let
\begin{eqnarray}
\label{bayesian-crude}
\widehat{\Theta}({\bf x}_1^{1:N},{\bf x}_2^{1:N})
& = &
\sum_{i=1}^N w^i({\bf x}_1^{1:N}) f(x_1^i,x_2^i), \\
\label{bayesian-CMC}
\widetilde{\Theta}({\bf x}_1^{1:N})
& = &
\sum_{i=1}^N w^i({\bf x}_1^{1:N}) 
\left[ \int f(x_1^i,x_2) p(x_2|x_1^i) {\rm d}x_2 \right].
\end{eqnarray}
In this paper, 
we shall call 
$\widehat{\Theta}({\bf x}_1^{1:N},{\bf x}_2^{1:N})$ (resp. 
$\widetilde{\Theta}({\bf x}_1^{1:N})$)
the Bayesian crude MC 
(resp. Bayesian CMC) estimator of $\Theta$. 

\subsubsection{Discussion}
\label{temporal-spatial-discussion}

Let us now compare $\widetilde{\Theta}$ to $\widehat{\Theta}$.
As in section \ref{intro-variance-reduction},
$\widetilde{\Theta}$ outperforms 
$\widehat{\Theta}$ for all $N$,
but not for the same reasons.
Indeed we have 
\begin{equation} 
\label{egalite-esperance}
{\rm E}(w^i({\bf X}_1^{1:N}) 
f(X_1^i,X_2^i)|{\bf x}_1^{1:N})= 
w^i({\bf x}_1^{1:N}) 
\int f(x_1^i,x_2) p(x_2|x_1^i) {\rm d}x_2 .
\end{equation} 
So from \eqref{inegalite-variances-RB},
the variance of each term of 
\eqref{bayesian-CMC} is lower than or equal to 
that of the corresponding term in 
\eqref{bayesian-crude};
however 
this is not sufficient
to conclude that 
${\rm var }(\widetilde{\Theta}) \leq {\rm var} (\widehat{\Theta})$
since the terms may be dependent.
Fortunately 
\eqref{egalite-esperance}
implies that 
$\widetilde{\Theta}={\rm E}(\widehat{\Theta} | {\bf x}_1^{1:N} )$,
so 
$\widetilde{\Theta}$ is preferable to 
$\widehat{\Theta}$,
due to 
\eqref{egalite-esperances-conditionnel} and
\eqref{inegalite-variances-RB}.

Let us now turn to practical considerations.
Of course,
$\widetilde{\Theta}$ is of interest 
only if the conditional expectation
in \eqref{bayesian-CMC}
can be computed easily.
In the rest of this paper
we will see that this indeed is the case
in some Markovian models and for other models,
we will propose and discuss some approximations
which make the Bayesian CMC estimator
a tool of practical interest for practitioners  
which may be used as an alternative 
to purely Monte Carlo classical PF. 
From a modeling point of view,
by contrast with spatially partitioned RB-PF,
the state space no longer needs to be multi-dimensional;
here a key point is the availability 
(and integrability) of $p(x_2|x_1)$,
which,
in the temporal partitions considered below,
will coincide with the so-called optimal conditional importance distribution.
From a numerical point of view, 
another interesting feature of sequential RB-PF 
is that numerical approximations,
when necessary,
do not need to be propagated over time.

Let us finally address complexity.
As we shall see,
in some cases 
$\widetilde{\Theta}$ 
can even be computed 
under the same assumptions
and for the same computational cost
as $\widehat{\Theta}$
(see sections 
\ref{exact-computation-CMC-hmc} and
\ref{practice-jmss1}).
Also one should note that in the partition $(X_1,X_2)$ of a given set of variables
(${\bf X}_{0:n}$, say)
$X_1$ should be as small as possible.
More precisely,
let 
$\Theta = 
{\rm E}
(f(X_1,X_2,X_3))
$
and let
$\widehat{p}(x_1,x_2)$ be available.
Then two Bayesian CMC estimators can be thought of :
$\widetilde{\Theta}^{X_3}$
built from 
$
\Theta 
=
{\rm E}
[
{\rm E}
(f(X_1,X_2,X_3)|X_1,X_2)
]
$,
in which 
the inner expectation (w.r.t. $X_3$)
is computed exactly, 
and
$\widetilde{\Theta}^{(X_2,X_3)}$ 
built from
$
\Theta 
=
{\rm E}
[
{\rm E}
(f(X_1,X_2,X_3)|X_1)
]
$
and from $\widehat{p}(x_1)$.
Estimator $\widetilde{\Theta}^{(X_2,X_3)}$ is preferable to
$\widetilde{\Theta}^{X_3}$,
but computing $\widetilde{\Theta}^{(X_2,X_3)}$ 
requires an additional exact expectation computation,
since
$
{\rm E}
(f(X_1,X_2,X_3)|X_1)
=
{\rm E}
[
{\rm E}
(f(X_1,X_2,X_3)|X_1,X_2)
]
$.
As we shall see in section \ref{practice-jmss1},
in some Markovian models
both estimators can indeed be computed;
and computing $\widetilde{\Theta}^{(X_2,X_3)}$ 
only requires an additional computational cost.

The rest of this paper is organized as follows.
First in section
\ref{semi-exact-hmc}
we see that in some HMC models
(including the Autoregressive Conditional Heteroscedasticity (ARCH) ones),
a Bayesian CMC estimator $\widetilde{\Theta}_n$
can replace the classical one $\widehat{\Theta}_n$
in the case where 
the sampling importance resampling (SIR) algorithm
with optimal importance distribution
is used.
%
%
In Section \ref{jmss}
we develop our Bayesian CMC estimators 
for JMSS;
in section \ref{linear-gaussian-jmss}
we address the linear and Gaussian case,
where our solution can be seen as a further (temporal) RB step
of an already (spatial) RB-PF algorithm;
in section \ref{extension-jmss}
we develop Bayesian CMC estimators for general JMSS.
Finally in Section \ref{multi-phd}
we address a multi-target scenario
and adapt Bayesian CMC
to the 
PHD filter.
In all these sections 
we propose relevant approximate estimators 
when the Bayesian CMC estimator cannot be computed exactly,
and we validate our algorithms via simulations.
We finally end the paper with a Conclusion.

\section{Bayesian CMC PF for some HMC models}
\label{semi-exact-hmc}

\subsection{Deriving a Bayesian CMC estimator $\widetilde{\Theta}_n$}
\label{deriving-hmc}

Let 
$\{{\bf X}_n\}_{n\geq 0}$ (resp. 
$\{{\bf Y}_n\}_{n\geq 0}$)
be a 
$p$- (resp. 
$q$-) dimensional state vector
(resp. observation).
We assume that $({\bf X}_n,{\bf Y}_n)$
follows the well known HMC model:
\begin{equation}
\label{hmc}
p({\bf x}_{0:n},{\bf y}_{0:n}) =
p({\bf x}_{0})
\prod_{i=1}^n
f_{i|i-1}({\bf x}_{i}|{\bf x}_{i-1})
\prod_{i=0}^n
g_i({\bf y}_{i}|{\bf x}_{i}),
\end{equation}
in which 
$f_{i|i-1}(\x_i|\x_{i-1})$
is the transition pdf of Markov chain $\{ {\bf X}_n \}_{n \geq 0}$ and
$g_i({\bf y}_i|{\bf x}_i)$ the likelihood of ${\bf y}_i$ given ${\bf x}_i$.
The Bayesian filtering problem
consists in computing some moment of interest 
$\Theta_n = {\rm E}_{p_{n|n}}(f({\bf X}_n))$,
which we rewrite as 
\begin{eqnarray}
\label{def_thetan_FAAPF-b}
\Theta_n
& = &
\int
f(\x_n) p(\x_{0:n-1},\x_n|\y_{0:n}) {\rm d}\x_{0:n-1}  {\rm d}\x_n .
\end{eqnarray}
So
\eqref{def_thetan_FAAPF-b} coincides with 
\eqref{def-de-Theta-jointe},
with 
$X_1 = {\bf X}_{0:n-1}$,
$X_2 = {\bf X}_{n}$,
$f(x_1,x_2)$ depends on $x_2$ only, 
and 
$p(x_1,x_2)$ is the a posteriori 
(i.e., given $\y_{0:n}$)
joint pdf
\begin{equation}
\label{p_12-hmc}
p(\x_{0:n-1},\x_n|\y_{0:n})
=
\underbrace{
p(\x_{0:n-1}|\y_{0:n})
}_{p(x_1)}
\underbrace{
p(\x_{n}|\x_{n-1},\y_{n})
}_{p(x_2|x_1)}.
\end{equation}
According to \eqref{approx-p1}
we first need an approximation of $p(x_1)$, 
which in model \eqref{hmc} reads:
\begin{equation}
\label{p_n-1_n_hmc}
p(\x_{0:n-1}|\y_{0:n})
= 
\frac
{p(\y_{n}|\x_{n-1}) p(\x_{0:n-1}|\y_{0:n-1})}
{\int p(\y_{n}|\x_{n-1}) p(\x_{0:n-1}|\y_{0:n-1}) {\rm d}\x_{0:n-1}},
\end{equation}
in which
$p({\bf y}_n |{\bf x}_{n-1}) =$
$\int g_n({\bf y}_n |{\bf x}_n) f_{n|n-1}({\bf x}_n |{\bf x}_{n-1}) {\rm d}{\bf x}_n$.
On the other hand,
PF algorithm propagate approximations of 
$p_{0:n-1|n-1}$ or of
$p_{n-1|n-1}$.
So let us start from
$\widehat{p}(\x_{0:n-1}|\y_{0:n-1}) = 
\sum_{i=1}^N
w_{n-1}^i \delta_{\x_{0:n-1}^i}
$.
According to Rubin's SIR mechanism
\cite{rubin1988}
\cite{gelfand-smith}
\cite{smith-gelfand}
$\widehat{p}(\x_{0:n-1}|\y_{0:n}) = 
\sum_{i=1}^N
\widetilde{w}_{n-1}^i \delta_{\x_{0:n-1}^i}
$,
where
$
\widetilde{w}_{n-1}^i \propto
{w}_{n-1}^i 
p({\bf y}_n|{\bf x}_{n-1})
$,
is an approximation of 
$p(\x_{0:n-1}|\y_{0:n})$.
Next
$p(x_2|x_1)$ in
\eqref{p_12-hmc}
coincides with the so-called
optimal conditional importance pdf,
i.e. the importance density
$p({\bf x}_n |{\bf x}_{n-1}, {\bf y}_n) \propto$
$g_n({\bf y}_n |{\bf x}_n) f_{n|n} ({\bf x}_n | {\bf x}_{n-1})$
which minimizes the conditional variance of weights $w_n^i$,
given past trajectories and observations
\cite{zaritskii1975}
\cite{kong1994}
\cite{liu-chen1995} and
\cite{doucet-sequentialMC}.
This leads to the so-called SIR algorithm
with optimal importance distribution and 
optional resampling step:

\vspace{0.4cm}
\textbf{SIR algorithm.} 
Let 
$\widehat{p}_{0:n-1|n-1} = \sum_{i=1}^N 
w_{n-1}^i \delta_{\x_{0:n-1}^i}$
be an MC approximation 
of $p_{0:n-1|n-1}$.
\begin{enumerate}
\item 
For all $i$, 
$1\leq i \leq N$,
sample $\widetilde{\x}_n^i \sim p(\x_n|\x_{n-1}^i,\y_n)$;
\vspace{0.2cm}
\item 
For all $i$, 
$1\leq i \leq N$,
set $\widetilde{w}_n^i \propto w_{n-1}^i p(\y_n|\x_{n-1}^i)$, 
$\sum_{i=1}^N \widetilde{w}_n^i=1$;
\vspace{0.2cm}
\item (Optional). 
For all $i$, 
$1\leq i \leq N$,
(re)sample 
$\x_{0:n}^i \sim 
\sum_{i=1}^N 
\widetilde{w}_{n}^i 
\delta_{[{\bf x}_{0:n-1}^i,\widetilde{\x}_{n}^i]} $,
and set 
$w_n^i = \frac{1}{N}$;
otherwise
set 
$({\x}_{n}^i,{w}_n^i) = (\widetilde{\x}_{n}^i,\widetilde{w}_n^i)$.
\end{enumerate}
This third resampling step
is usually performed only if some criterion holds,
and aims at preventing weights degeneracy, 
see e.g.
\cite{livredoucet},
\cite{arulampalamshort}.
Then
\begin{equation}
\label{p-hat-SIR}
\widehat{p}_{0:n|n}^{\rm SIR} = 
\sum_{i=1}^N 
\widetilde{w}_{n}^i \delta_{\x_{0:n-1}^i,\widetilde{\x}_{n}^i}
\end{equation}
is a (SIR-based) SMC approximation of 
$p_{0:n|n}$, and
$\widehat{p}_{0:n-1,n|n}^{\rm SIR}$ 
plays the role of 
$\widehat{p}(x_1,x_2)$ in
\eqref{approx-p1p2}.
Finally from 
\eqref{bayesian-crude} and
\eqref{bayesian-CMC},
the SIR-based crude and CMC estimators 
of moment $\Theta_n$ defined in \eqref{def_thetan_FAAPF-b}
are respectively
\begin{eqnarray}
\label{bayesian-crude-sir}
\widehat{\Theta}_n^{\rm SIR}
(
{\bf x}_{0:n-1}^{1,N},
\widetilde{\bf x}_{n}^{1:N}
) 
& = &
\sum_{i=1}^N 
\widetilde{w}_n^i(
{\bf x}_{0:n-1}^{1:N}
)
f(\widetilde{\x}_{n}^i), \\
\label{bayesian-CMC-sir}
\widetilde{\Theta}_n^{\rm SIR}
(
{\bf x}_{0:n-1}^{1:N}
) 
& = &
\sum_{i=1}^N 
\widetilde{w}_n^i
(
{\bf x}_{0:n-1}^{1:N}
)
\int \!\!f({\bf x}_{n}) p({\bf x}_n|{\bf x}_{n-1}^i,{\bf y}_{n}) 
{\rm d}{\bf x}_{n} .
\end{eqnarray}

\subsection{Computing $\widetilde{\Theta}_n^{\rm SIR}$ in practice}
\label{computing-practice}

\subsubsection{Exact computation}
\label{exact-computation-CMC-hmc}

From \eqref{egalite-esperance}
we know that 
$\widetilde{\Theta}_n^{\rm SIR}$ outperforms
$\widehat{\Theta}_n^{\rm SIR}$;
but 
$\widetilde{\Theta}_n^{\rm SIR}$ can be used 
only if 
$\widetilde{w}_n^i$ and
integral 
$\int f({\bf x}_{n}) p({\bf x}_n|{\bf x}_{n-1}^i,{\bf y}_{n}) 
{\rm d}{\bf x}_{n}$
can be computed.
As we now see, 
this is the case in some particular HMC models
and for some functions $f({\bf .})$.
Let us e.g. consider the semi-linear
stochastic models with additive Gaussian noise,
given by 
\begin{eqnarray}
\label{semi-etat}
{\bf X}_{n}&=&{\rm \bf f}_n({\bf X}_{n-1}) + {\rm \bf K}_n({\bf X}_{n-1}) \times {\bf U}_n , \\
\label{semi-observation}
{\bf Y}_{n}&=&{\rm \bf H}_n  {\bf X}_n + {\bf V}_n ,
\end{eqnarray}
in which $\{{\bf U}_n\}$ and $\{{\bf V}_n\}$ are i.i.d.,
mutually independent and
independent of ${\bf X}_0$,
${\bf U}_n \sim \mathcal{N}(0;{\bf I})$ and 
${\bf V}_n \sim \mathcal{N}(0;{\bf R}_n^v)$.
The one-dimensional ARCH model
is one such model with ${ f}_n(x_{n-1})=0$,
${k}_n(x_{n-1})= \sqrt{\beta_0+\beta_1x_{n-1}^2}$ and
${H}_n=1$.  
In 
model
\eqref{semi-etat}
\eqref{semi-observation}
$p(\x_n|\x_{n-1},\y_n)$ and
$p(\y_n|\x_{n-1})$ are Gaussian.
More precisely,
let
${\bf Q}_n(\x_{n-1})={\rm \bf K}_n(\x_{n-1}) {\rm \bf K}_n(\x_{n-1})^T$;
then
\begin{eqnarray}
\label{Ln}
{\bf L}_n(\x_{n-1})
& = &
{\bf H}_n{\bf Q}_n(\x_{n-1}){\bf H}_n^T+{\bf R}_n^v,
\\
{\bf m}_n(\x_{n-1},\y_n)& = &{\bf f}_n(\x_{n-1})+{\bf Q}_n(\x_{n-1}) {\bf H}_n^T
{\bf L}_n^{-1}(\x_{n-1}) 
({\bf y}_{n}-{\bf H}_n {\bf f}_n(\x_{n-1}))
\text{,}  
\\
{\bf P}_n(\x_{n-1})
& = &
{\bf Q}_n(\x_{n-1})
-
{\bf Q}_n(\x_{n-1}){\bf H}_n^T
{\bf L}_n^{-1}(\x_{n-1})  
{\bf H}_n
{\bf Q}_n(\x_{n-1})
\text{,}  
\\
\label{sampling-distrib}
p(\x_n|\x_{n-1},\y_n) 
& = & 
\mathcal{N}(\x_n,{\bf m}_n(\x_{n-1},\y_n),{\bf P}_n(\x_{n-1})) 
\text{,}  
\\
\label{semi-poids}
p(\y_n|\x_{n-1})
& = & 
\mathcal{N}(\y_n,{\bf H}_n {\bf f_n}(\x_{n-1}),
{\bf L}_n(\x_{n-1}) 
) \text{.} 
\end{eqnarray}
Finally in such models the Bayesian CMC estimator
$\widetilde{\Theta}_n^{\rm SIR}$
is workable for some functions $f({\bf .})$.
If $f(\x)$
is a polynomial in $\x$, 
the problem
reduces to computing the first moments 
of the available Gaussian pdf \eqref{sampling-distrib}.
In the important particular case where
$f(\x)=\x$
(used to give an estimator
of the hidden state),
no further
computation is indeed necessary;
in this case the integral in 
\eqref{bayesian-CMC-sir}
is equal to ${\bf m}_n(\x_{n-1}^i,\y_n)$.

\begin{remark}
In this class of models, computing 
$\widetilde{\Theta}_n^{\rm SIR}$
or 
$\widehat{\Theta}_n^{\rm SIR}$
requires the same computational cost
if $f({\bf x}) = {\bf x}$.
Both estimators indeed compute 
the parameters
${\bf m}_n(\x_{n-1}^i,\y_n)$ and
${\bf P}_n(\x_{n-1}^i)$
of $p(\x_n|\x_{n-1}^i,\y_n)$, 
and use these pdfs 
to sample the new particles $\widetilde{\bf x}_n^i$,
which in both cases are needed for the next time step.
The only difference is that 
$\widehat{\Theta}_n^{\rm SIR} =$
$\sum_{i=1}^N \widetilde{w}_n^i \widetilde{\x}_{n}^i$,
while 
$\widetilde{\Theta}_n^{\rm SIR} =$
$\sum_{i=1}^N \widetilde{w}_n^i 
{\bf m}_n(\x_{n-1}^i,\y_n)$.

\end{remark}

\subsubsection{Approximate computation}
\label{approximate-computation-CMC-hmc} 

Let us now discuss cases 
where the Bayesian CMC estimator
$\widetilde{\Theta}_n^{\rm SIR}$ cannot be computed exactly because 
$p(\y_n|\x_{n-1})$ and/or moments of
$p(\x_n|\x_{n-1},\y_n)$ 
are not computable.
Two approximations are proposed:

\begin{itemize}
\item 
Available techniques such as local
linearizations \cite{doucet-sequentialMC},
Taylor series expansion \cite{Saha_EMM}
or the Unscented Transformation (UT) \cite{julier-procieee} 
have already been proposed
for approximating $p(\y_n|\x_{n-1})$ and a moment of
$p(\x_n|\x_{n-1},\y_n)$,
so one can use any of them in \eqref{bayesian-CMC-sir}.
The resulting algorithm can be seen as an alternative
to solutions like the Extended Kalman Filter (EKF) or 
the Unscented Kalman filter (UKF),
where we look for approximating the filtering pdf $p_{n|n}$
by a Gaussian 
and which rely on linearizations or the UT;
or to SMC methods,
where we look for a discrete approximation of $p_{n|n}$. 
In our approximate Bayesian CMC technique,
we start from a discrete approximation of $p_{n-1|n-1}$
produced by an SMC method,
then similarly to the EKF/UKF,
we look for a numerical approximation
of $\widetilde{\Theta}_n$,
given that discrete approximation of $p_{n-1|n-1}$.
However, deriving a good approximation of $p(\y_n|\x_{n-1})$
can be an intricate issue,
so we next look for approximations
which do not rely on an approximation
of $p(\y_n|\x_{n-1})$.

\item 
In the SIR algorithm used so far,
$\widetilde{\x}_n^i$ is drawn 
from 
$p(\x_n|\x_{n-1}^i,\y_n)$,
whence a weight update factor equal to $p(\y_n|\x_{n-1}^i)$.
On the other hand, 
sampling $\widetilde{\x}_{n}^i$
from an alternate (i.e., not necessarily optimal) pdf 
$q(\x_n|\x_{n-1}^i)$
yields an approximation of $p_{0:n-1|n}$ 
given by 
$\widehat{p}_{0:n-1|n} =
\sum \widetilde{w}_{n}^i \delta_{\x_{0:n-1}^i}$,
where weights $\widetilde{w}_{n}^i$ are now proportional to
$
w_{n-1}^i
f_{n|n-1}(\widetilde{\bf x}_n^i|\x_{n-1}^i)
g_n(\y_n|\widetilde{\bf x}_n^i)
/
q(\widetilde{\bf x}_n^i|\x_{n-1}^i)$, 
and so depend also on the new samples 
$\{\widetilde{\bf x}_n^i\}_{i=1}^N$. 
In that case, 
the associated Bayesian CMC and crude estimators 
become
\begin{eqnarray}
\label{crude-sir-quelconque}
\widehat{\Theta}_n
(
\x_{0:n-1}^{1:N},\widetilde{\bf x}_{n}^{1:N}
) 
& \!\!= \!\!&
\sum_{i=1}^N 
\widetilde{w}_n^i
(
\x_{0:n-1}^{1:N},\widetilde{\bf x}_{n}^{1:N}
) 
f(\widetilde{\bf x}_n^i),
\\
\label{cmc-sir-quelconque}
\widetilde{\Theta}_n
(
\x_{0:n-1}^{1:N},\widetilde{\bf x}_{n}^{1:N}
) 
& \!\!= \!\!&
\sum_{i=1}^N \widetilde{w}_n^i   
(
\x_{0:n-1}^{1:N},\widetilde{\bf x}_{n}^{1:N}
) 
\!\!\!
\int 
\!\! f({\bf x}_{n}) p({\bf x}_n|{\bf x}_{n-1}^i,{\bf y}_{n}) {\rm d}\x_n,
\end{eqnarray}
which 
can no longer be compared easily
(it was the case in section \ref{bayesian-cmc-estimators},
because the weights $w^i$ in
$\widehat{\Theta}$   in \eqref{bayesian-crude} and
$\widetilde{\Theta}$ in \eqref{bayesian-CMC}
depend on $\{x_1^i\}_{i=1}^N$ only).
On the other hand,
the computation of 
\eqref{cmc-sir-quelconque}
does not require 
that of $p(\y_n|\x_{n-1})$,
but only that of 
$\int f({\bf x}_{n}) p({\bf x}_n|{\bf x}_{n-1}^i,{\bf y}_{n}){\rm d} \x_n$.
This is of interest in some models
where
approximating 
$p(\y_n|\x_{n-1})$ 
may be challenging because of the form of $g_n$, 
while the first order moments of $p(x_n|x_{n-1},y_n)$ 
can be computed or approximated easily \cite{Saha_EMM}.
\end{itemize} 
The two approximate implementations of the Bayesian CMC estimator 
which we just discussed
will be compared via simulations in section 
\S \ref{simu-SV}.

\subsection{Alternate Bayesian CMC solutions}
\label{alternate-CMC}

\subsubsection{A Bayesian CMC estimator based
on the fully-adapted auxiliary particle filter (FA)}
\label{CMC-FA}

The SIR algorithm of section \ref{deriving-hmc}
is not the only SMC algorithm 
which enables to compute 
an approximation of $p(\x_{0:n-1},\x_n|\y_{0:n})$
in which weights 
depend on $\{ \x_{0:n-1}^i \}_{i=1}^N$ only.
Starting from 
$\widehat{p}(\x_{0:n-1}|\y_{0:n-1}) = 
\sum_{i=1}^N
w_{n-1}^i \delta_{\x_{0:n-1}^i}$,
the so-called FA algorithm 
\cite{auxiliary} 
\cite{fearnhead}
is one such alternative:

\vspace{0.4cm}
\textbf{FA algorithm.} 
Let 
$\widehat{p}_{0:n-1|n-1} = \sum_{i=1}^N 
w_{n-1}^i \delta_{\x_{0:n-1}^i}$
be an MC approximation 
of $p_{0:n-1|n-1}$.
\begin{enumerate}
\item 
For all $i$, 
$1\leq i \leq N$,
set $\widetilde{w}_n^i \propto w_{n-1}^i p(\y_n|\x_{0:n-1}^i)$, $\sum_{i=1}^N \widetilde{w}_n^i=1$;
\vspace{0.2cm}
\item 
For all $i$, 
$1\leq i \leq N$,
sample $\widetilde{\x}_{0:n-1}^i \sim \sum_{i=1}^N 
\widetilde{w}_{n}^i 
\delta_{\x_{0:n-1}^i} $,
\vspace{0.2cm}
\item 
For all $i$, 
$1\leq i \leq N$,
sample 
$\x_n^i \sim p(\x_n|\widetilde{\x}_{n-1}^i,\y_n)$ 
and set 
$w_n^i=\frac{1}{N}$,
${\bf x}_{0:n}^i =
[\widetilde{\x}_{0:n-1}^i,{\bf x}_n^i]$.
\end{enumerate}
Finally 
\begin{equation}
\label{p-hat-FA}
\widehat{p}_{0:n-1,n|n}^{\rm FA} = 
\sum_{i=1}^N 
\frac{1}{N} \delta_{\widetilde{\x}_{0:n-1}^i,\x_n^i}
\end{equation}
is the FA-based SMC approximation of 
$p_{0:n-1,n|n}$, 
and the FA-based crude and CMC estimators of $\Theta_n$
become respectively 
\begin{eqnarray}
\label{bayesian-crude-FA}
\widehat{\Theta}_n^{\rm FA}
(
\widetilde{\bf x}_{0:n-1}^{1:N},
{\bf x}_{n}^{1:N}
)
& = &
\widehat{\Theta}_n^{\rm FA}
(
{\bf x}_{n}^{1:N}
)
=
\frac{1}{N}
\sum_{i=1}^N
f({\x}_{n}^i), \\
\label{bayesian-CMC-FA}
\widetilde{\Theta}_n^{\rm FA}
(
\widetilde{\bf x}_{0:n-1}^{1:N}
)
& = &
\widetilde{\Theta}_n^{\rm FA}
(
\widetilde{\bf x}_{n-1}^{1:N}
) 
=
\frac{1}{N}
\sum_{i=1}^N 
\int f({\bf x}_{n}) p({\bf x}_n|\widetilde{\bf x}_{n-1}^i,{\bf y}_{n}) 
{\rm d}{\bf x}_{n} .
\end{eqnarray}

\subsubsection{Discussion}
\label{discussion-FA}

Comparing with section \ref{deriving-hmc},
we see that two Bayesian CMC estimators 
are indeed available: 
the SIR-based one 
$\widetilde{\Theta}_n^{\rm SIR}$ 
given by 
\eqref{bayesian-CMC-sir},
and the FA-based one 
$\widetilde{\Theta}_n^{\rm FA}$ 
given by \eqref{bayesian-CMC-FA}.
The natural question which arises at this point
is thus to wonder which one is best.
Two arguments are available.

Let us first start 
from a common MC approximation
$\widehat{p}_{0:n-1|n-1} =$ $ \sum_{i=1}^N$
$w_{n-1}^i \delta_{\x_{0:n-1}^i}$ of $p_{0:n-1|n-1}$.
Given $\widehat{p}_{0:n-1|n-1}$ and ${\bf y}_n$,
trajectories $\{\widetilde{\x}_{0:n-1}^i \}_{i=1}^N$ 
produced by the FA algorithm 
are i.i.d. from 
$\sum_{i=1}^N 
\widetilde{w}_{n}^i 
\delta_{\x_{0:n-1}^i}$.
As is well known,
resampling introduces variance, 
so given $\hat{p}_{0:n-1|n-1}$
$\widetilde{\Theta}_n^{\rm SIR}$ is preferable to 
$\widetilde{\Theta}_n^{\rm FA}$, and
$\widetilde{\Theta}_n^{\rm FA}$ should not be used in practice.

On the other hand,
the performances of
$\widetilde{\Theta}_n^{\rm SIR}$ 
also depend on the weighted trajectories
$\{( {\x}_{0:n-1}^i, w_{n-1}^i ) \}_{i=1}^N$
which are available at time $n-1$;
so one can wonder whether one should 
propagate them via the SIR algorithm,
or via the FA one.

This actually is a thorny issue,
because in the SIR algorithm
the resampling step is optional 
and is often performed according to a particular criterion, 
like an estimator of the so-called number of efficient particles 
\cite{kong1994} \cite{liu-chen1995}.
So comparing the set 
$\{ \widetilde{\x}_{n}^i, \widetilde{w}_{n}^i \}_{i=1}^N$
produced by the SIR algorithm \textit{before} 
the resampling step 
to that 
$\{ \x_{n}^i, 1/N 
\}_{i=1}^N$
produced by the FA algorithm, 
is a challenging task, 
and indeed it has been proved in \cite{doucet-APF} 
(from an asymptotical point of view)
that none algorithm always outperforms the other.

If however we assume that the resampling step 
is done at each time step, 
then 
it is well known \cite[Ch. 9]{Cappeetal}
that the set of samples produced by 
the  FA algorithm is better (in an asymptotic normality sense) 
than that produced by the SIR algorithm
\textit{after} the resampling step.
This can be easily understood empirically 
from a simple argument.
Starting from a set of weighted samples 
$\{ \x_{0:n-1}^i , w_{n-1}^i \}_{i=1}^N$,
the number of different particles 
$\{\x_{n}^i\}$ 
produced by the FA algorithm 
is equal to $N$,
while that produced by the SIR one (after resampling) 
is lower than $N$,
and can consequently lead to a poor approximation of $p_{n|n}$. 

\subsection{Simulations}
\label{simu-hmc}
In section \ref{Gaussian-simu}
we compare via simulations
two Bayesian CMC estimators 
$\widetilde{\Theta}_n^{\rm SIR}$,
which differ only by the set of weighted points $\widehat{p}_{0:n-1|n-1}$
upon which they rely at each time instant $n-1$:
this set will be either propagated by the SIR algorithm 
($\widetilde{\Theta}_n^{{\rm SIR},1}$),
or by the FA one
($\widetilde{\Theta}_n^{{\rm SIR},2}$).
In section \ref{arch-simu} we compare
$\widetilde{\Theta}_n^{\rm FA}$ and
$\widetilde{\Theta}_n^{{\rm SIR},2}$ 
in the ARCH model.
In section \ref{simu-SV} 
we compare the two approximations
of $\widetilde{\Theta}_n$ 
described in section \ref{approximate-computation-CMC-hmc},
and the weighted trajectories 
are propagated by the SIR algorithm.
We compute the empirical mean square 
error (MSE) at each time step, 
averaged on $P=200$ simulations;
the true mean is computed by the Kalman filter (KF) in the Gaussian case,
or a bootstrap filter \cite{gordon-salmond-smith}
with $N=10^5$ particles otherwise.

\subsubsection {Gaussian Model}
\label{Gaussian-simu}
\begin{figure}[htbp]
\center
\epsfig{file=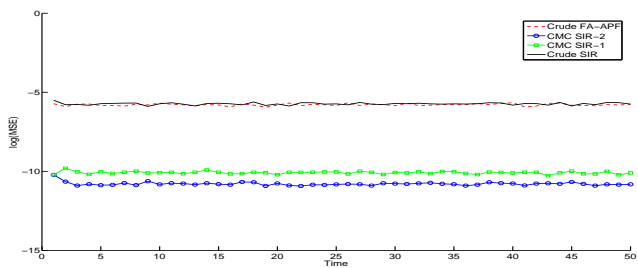,height=0.2\textheight,width=0.85\textwidth} 
\caption{\label{fig-MSE-GAUSS} 
MSE - Gaussian model, $R=1$, $Q=10$, $N=1000$ - $f(x_n)=x_n$. Estimator CMC-SIR-2 which
propagates the samples with an FA algorithm outperforms CMC-SIR-1 which uses a SIR algorithm. 
Both CMC estimators outperform the crude ones.
}
\end{figure}

We first consider a linear and Gaussian model
described by \eqref{semi-etat}-\eqref{semi-observation}
where $f_n(x_{n-1})=0.9x_{n-1}$, $H_n=1$,
$k_n(x_{n-1})=\sqrt{10}$ and $R_n^v=1$. 
We want to estimate the hidden state,
so $f(x_n)=x_n$. 
We compute the SIR- and FA-based 
Bayesian crude and CMC estimators with $N=1000$ particles;
of course KF, which computes ${\rm E}_{p_{n|n}}(x)$ exactly,
is here the benchmark solution.
MSEs of the four estimators are 
displayed in Fig. \ref{fig-MSE-GAUSS}.
$\widetilde{\Theta}_n^{{\rm SIR},1}$ (resp. 
$\widetilde{\Theta}_n^{{\rm SIR},2}$)
always outperforms
$\widehat{\Theta}_n^{\rm SIR}$ (resp.
$\widehat{\Theta}_n^{\rm FA}$).
Note also that
$\widehat{\Theta}_n^{\rm FA}$ does not always outperform
$\widehat{\Theta}_n^{\rm SIR}$,
which is in accordance
with the asymptotical analysis \cite{doucet-APF};
while
$\widetilde{\Theta}_n^{{\rm SIR},2}$ always outperforms 
$\widetilde{\Theta}_n^{{\rm SIR},1}$.

\subsubsection {ARCH Model}
\label{arch-simu}
\begin{figure} [htbp]
\center
	\epsfig{file=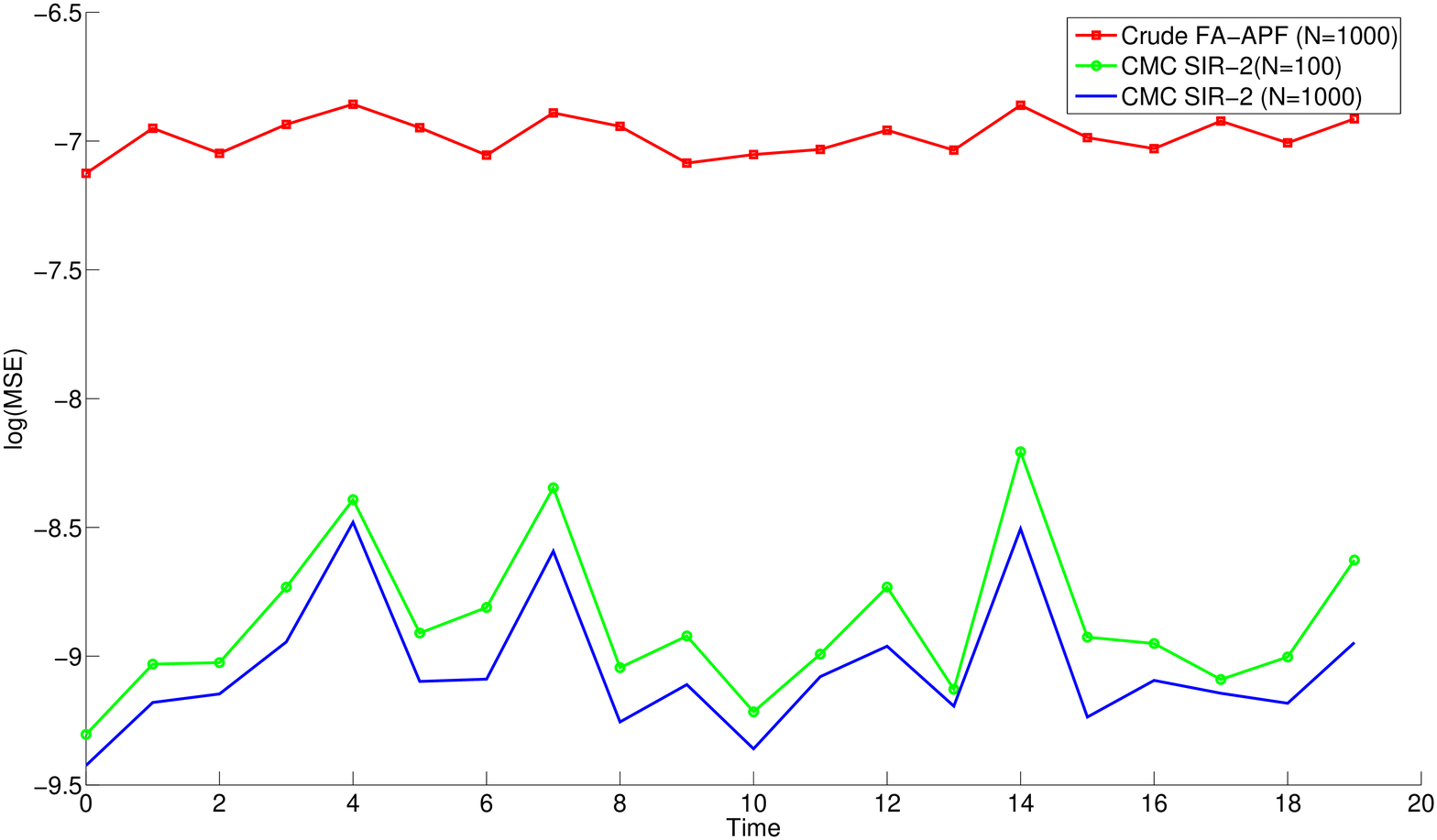,height=0.2\textheight,width=0.85\textwidth} 
	\caption{MSE - ARCH Model - $\beta_0=1$, $\beta_1=0.1$ and  $R_n^v=3$ - $f(x_n)=x_n$. The CMC estimator with $N=100$ particles
	outperforms the crude one with $N=1000$ particles and is close to the CMC one with $N=1000$ particles. }
	\label{arch-fig-x}
\end{figure}

\begin{figure} [htbp]
\center
\epsfig{file=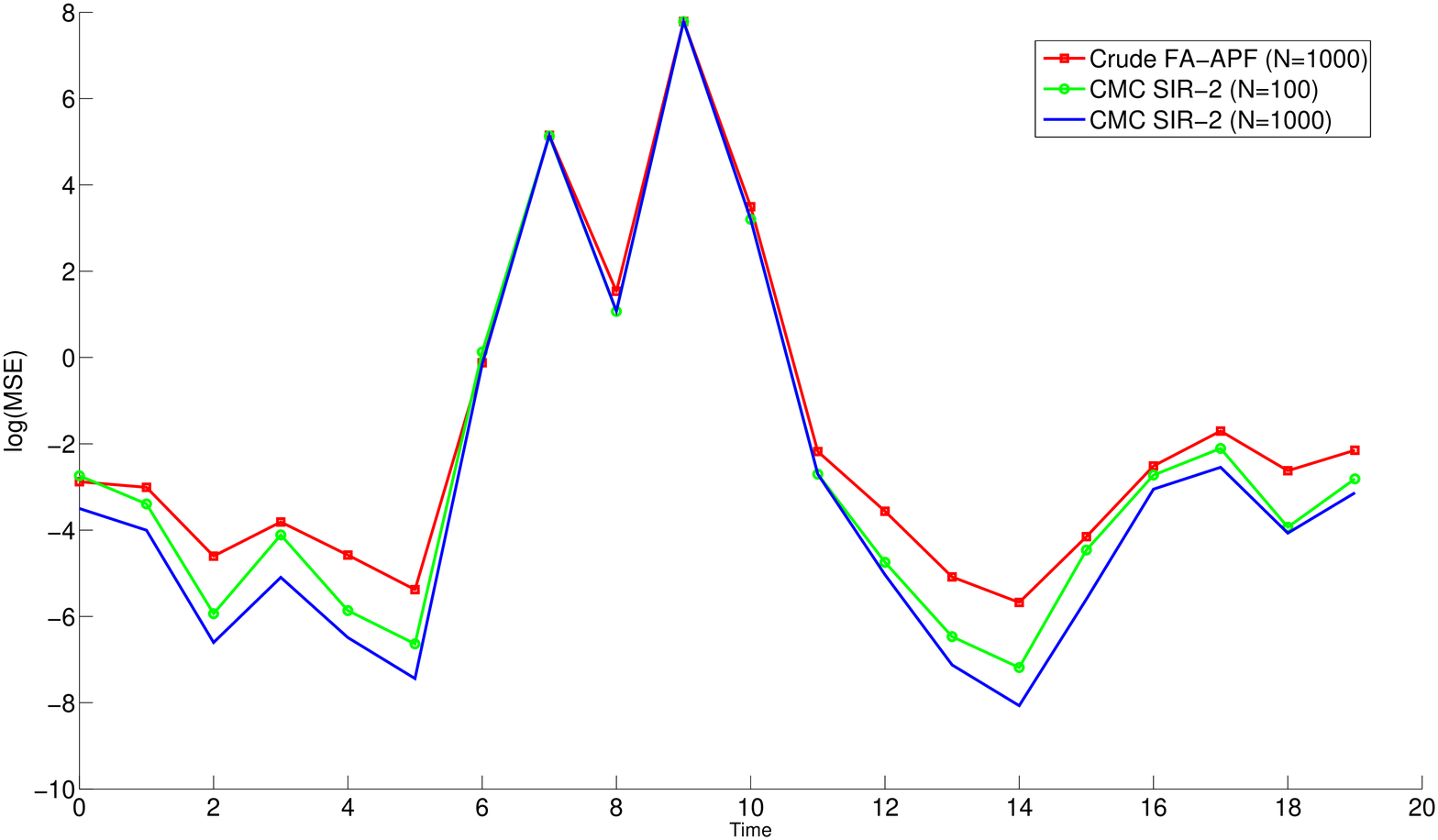,height=0.2\textheight,width=0.85\textwidth} 
	\caption{MSE - ARCH Model - $\beta_0=1$, $\beta_1=0.1$ and  $R_n^v=3$ - $f(x_n)=\beta_0+\beta_1 x_{n-1}^2$. The gap between the three estimators depends on function $f(.)$ and has decreased compared to the previous simulation. }
	\label{arch-fig}
\end{figure}

We next consider the ARCH model 
recalled in section \ref{exact-computation-CMC-hmc}
We set $R_n^v=3$, $\beta_0=1$ and $\beta_1=0.1$.
We want to estimate 
$x_n$ (so $f(x_n)=x_n$), 
and the variance of the process noise 
(so $f(x_n)=\beta_0+\beta_1 x_n^2)$.
Since $p(x_n|x_{n-1},y_n)$ is Gaussian
(see \eqref{sampling-distrib}),
it is possible to calculate both moments.
We compare
$\widehat{\Theta}_n^{\rm FA}(1000)$ and
$\widetilde{\Theta}_n^{{\rm SIR},2}(1000)$,
both computed with $N = 1000$ particles, and
$\widetilde{\Theta}_n^{{\rm SIR},2}(100)$. 
MSEs are displayed on Fig. \ref{arch-fig-x} for the estimate of $x_n$ 
and Fig. \ref{arch-fig} for the variance of the process noise.
As we see 
$\widetilde{\Theta}_n^{{\rm SIR},2}(1000)$, and even
$\widetilde{\Theta}_n^{{\rm SIR},2}(100)$,
both outperform
$\widehat{\Theta}_n^{\rm FA}(1000)$.
However the gap between the three algorithms 
is function dependent
and so the previous considerations are model and function dependent.

\subsubsection {Stochastic Volatility Model}
\label{simu-SV}
Let us consider the following model:
\begin{eqnarray}
\label{stochastic-volatility}
X_{n+1} =& \Phi X_n + U_n \\
Y_n =& \beta {\rm exp}(X_n/2) \times V_n
\end{eqnarray}
in which 
$U_n \sim \mathcal{N}(0,\sigma^2)$ and 
$V_n \sim \mathcal{N}(0,1)$.
In this model $\widetilde{\Theta}_n$ is not computable, 
whatever function $f$,
because $p(y_n|x_{n-1})$ is not computable.
We propose to compare 
two approximations of the Bayesian CMC estimator 
with a SIR based crude estimator.
Our first approximation $\overline{\Theta}_n^{{\rm SIR},1}$ only 
relies on the approximation of $\int f(x_n)p(x_n|x_{n-1},y_n) {\rm d}x_n$
(second item in \S \ref{approximate-computation-CMC-hmc})
while the second one $\overline{\Theta}_n^{{\rm SIR},2}$
relies in addition on that of $p(y_n|x_{n-1})$
(first item in \S \ref{approximate-computation-CMC-hmc}).
In this model, an approximation of 
$p(y_n|x_{n-1})$ is obtained by a
first order Taylor series expansion of function
${\rm log}(g_n(y_n|x_n))$ in $\Phi x_{n-1}$.
If the deduced approximation of 
$g_n(y_n|x_n)$ is noted 
$\widehat{g}_n(y_n|x_n)$ then
$\widehat{p}(y_n|x_{n-1})=\int f_{n|n-1}(x_n|x_{n-1}) 
\widehat{g}_n(y_n|x_n) {\rm d} x_n$
where 
$f_{n|n-1}(x_n|x_{n-1})=
\mathcal{N}(x_n;\Phi x_{n-1}; \sigma^2)$,
is now computable. 
If $\sigma$ is small, 
$f_{n|n-1}(x_n|x_{n-1})$ is approximately non-null
for values close to $\Phi \x_{n-1}$, 
and for such values $\widehat{g}_n(y_n|x_n)$
is a good approximation of $g_n(y_n|x_n)$. 
So one should get a good
approximation $\widehat{p}(y_n|x_{n-1})$
when $\sigma$ is small.
Finally, a deduced approximation
of $p(x_n|x_{n-1},y_n)$ is 
given by a Gaussian pdf, see \cite{auxiliary}.

\begin{figure} [htbp]
\center
\epsfig{file=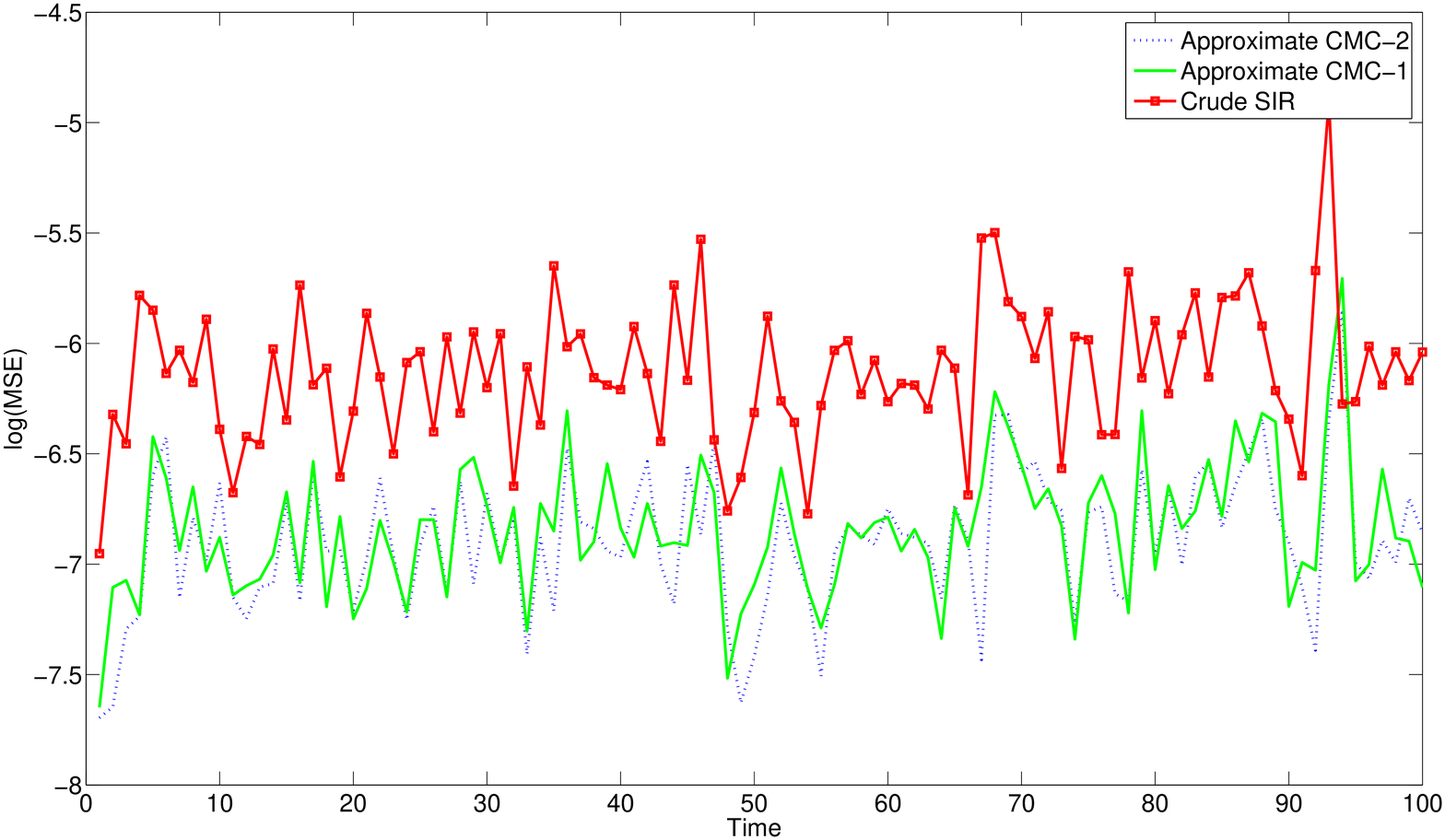,height=0.2555\textheight,width=0.85\textwidth}
	\caption{MSE - Stochastic Volatility Model - $\Phi=0.8$, $\beta=0.6$, $\sigma=0.18$, $N=1000$. Both
	approximate CMC estimators outperform the crude one. }
	\label{figure-SV}
\end{figure}

\begin{figure} [htbp]
\center
\epsfig{file=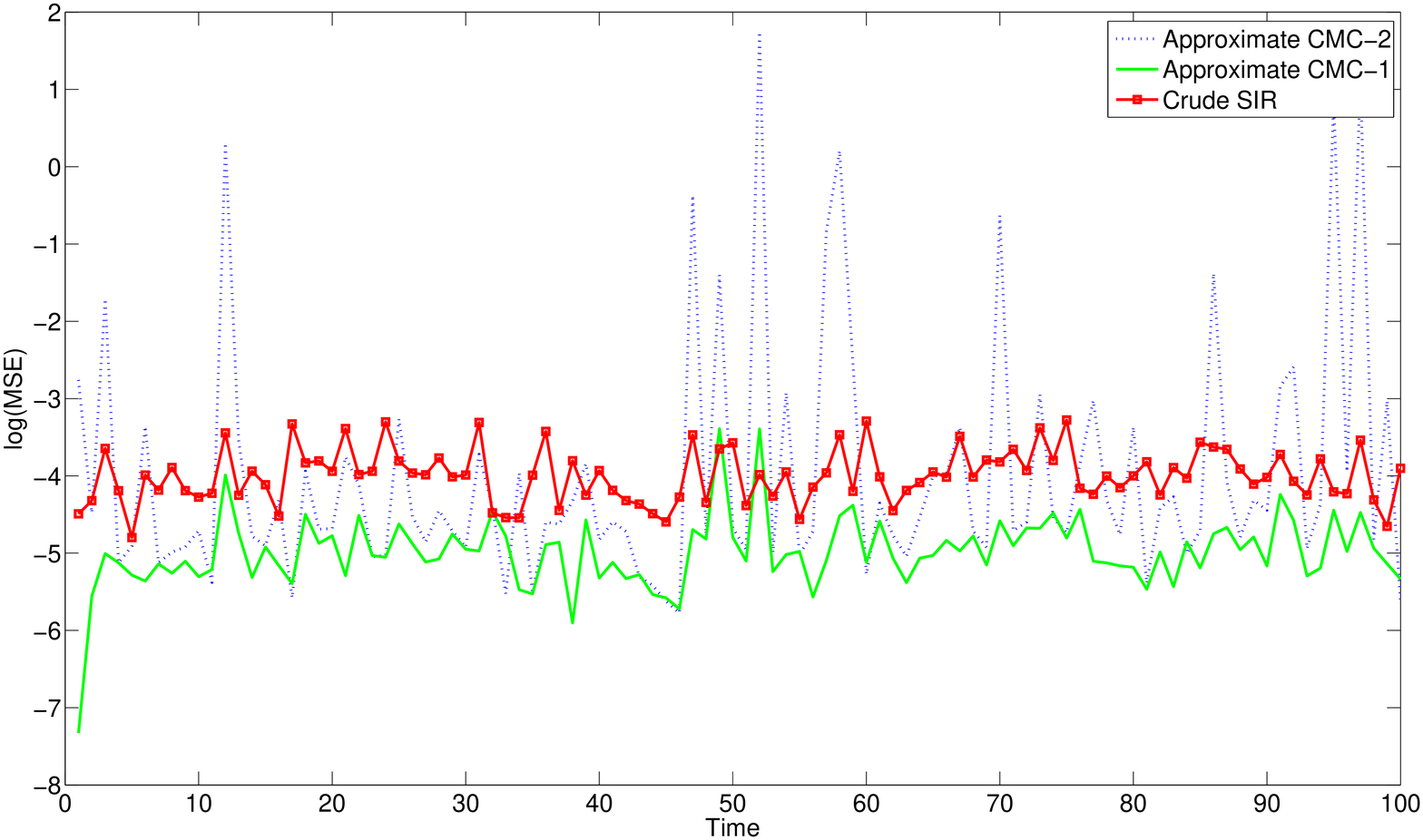,height=0.2555\textheight,width=0.85\textwidth} 
	\caption{MSE - Stochastic Volatility Model - $\Phi=0.8$, $\beta=0.6$, $\sigma=0.4$, $N=1000$.
Only the approximate estimator CMC-1 outperforms the crude estimator.
This is because the approximation of $p(\y_n|\x_{n-1})$ used 
by the approximate estimator CMC-2 is not reliable with these parameters. }
\label{figure-SV-2}
\end{figure}

We estimate the standard deviation
of the observation noise at time $n$
so $f(x_n)= \beta {\rm exp}( x_n/2)$.
We first take $\Phi=0.8$, $\beta=0.6$, $\sigma=0.18$.
Results are displayed in Fig. \ref{figure-SV}.
We observe that both approximations 
of the Bayesian CMC estimator 
outperform the crude SIR-based one, 
and that the second approximation, 
which does not use $\x_n^{1:N}$, is preferable.
However, in  Fig. \ref{figure-SV-2} we take
$\sigma=0.40$. Remember that increasing $\sigma$
has consequences on the approximation of $p(y_n|x_{n-1})$; 
as expected,
$\overline{\Theta}_n^{{\rm SIR},2}$,
which relies on this approximation,
is outperformed by the two other estimators. 
It is particularly interesting to notice that 
the first approximation $\overline{\Theta}_n^{{\rm SIR},1}$
is not affected and still outperforms the SIR based estimator.  
This confirms that Bayesian CMC estimators 
can still be of practical interest
in models which are not semi-linear.


\section {Bayesian CMC algorithms for JMSS models}
\label{jmss}
As in \S \ref{semi-exact-hmc}
we still consider the estimation of
$\Theta_n = \int \phi(\x_n) p(\x_n|\y_{0:n}) {\rm d}\x_n $
(the reason why we replaced $f$ by $\phi$
will become clear a few lines below),
but now in a so-called JMSS:
\begin{equation}
\label{hmc-jm}
p({\bf r}_{0:n},{\bf x}_{0:n},{\bf y}_{0:n}) \! = \!
p({ r}_{0}) \!
\prod_{i=1}^n \!
p({r}_{i}|{r}_{i-1})
p({\bf x}_{0}) \!
\prod_{i=1}^n \!
f_{i|i-1}({\bf x}_{i}|{\bf x}_{i-1},r_i) \!
\prod_{i=0}^n \!
g_i({\bf y}_{i}|{\bf x}_{i},r_i).
\end{equation}
Model \eqref{hmc-jm}
can be thought of as an HMC model,
in which
$f_{i|i-1}$ and
$g_i$
depend on the realization
of a discrete Markov Chain $\{ R_{n} \}_{n \geq 0}$
where each $R_n$
takes its values in $\{1, \cdots, K\}$.
So now both $X_n$ and $R_n$ are hidden,
and $\Theta_n$ can be rewritten as
\begin{equation}
\label{Theta_n-RB}
\Theta_n =
\sum_{{\bf r}_{0:n}}
\int
\phi({\bf x}_n) 
p({\bf x}_{n},{\bf r}_{0:n}|{\bf y}_{0:n})
{\rm d}{\bf x}_{n}.
\end{equation}
Note that $\phi(.)$ can also depend
on $r_n$. As is well known 
\cite{tugnait}
\cite{shalom-li-kirubarajan}
\cite{doucet-jump-Mkv},
in a JMSS exact Bayesian filtering is
either impossible (in the general case) or 
an NP-hard problem (in the linear and Gaussian case),
so one has to use suboptimal techniques.
Among them,
SMC methods can be divided into two classes:
\begin{itemize}
\item
In the first class 
\cite{musso-oudjane-legland}
\cite{ginnity}
\cite{Cappeetal}
$\Theta_n$ is computed
by injecting an SMC approximation of 
$p({\bf x}_{0:n},{\bf r}_{0:n}|{\bf y}_{0:n})$
into \eqref{Theta_n-RB};
\item
In the second class of SMC methods 
we start from 
\begin{equation}
\label{Theta-RB}
\Theta_n =
\sum_{{\bf r}_{0:n}}
\underbrace{
p({\bf r}_{0:n}|{\bf y}_{0:n})
}
_{\rm PF}
\int
\phi({\bf x}_n) 
\underbrace{
p({\bf x}_{n}|{\bf r}_{0:n},{\bf y}_{0:n})
}_
{\rm KF}
{\rm d}{\bf x}_{n} ,
\end{equation}
and propagate an SMC approximation 
$\sum_{i=1}^N w_n^i \delta_{{\bf r}_{0:n}^i}$
of
$p({\bf r}_{0:n}|{\bf y}_{0:n})$
only;
then
$\widehat{\Theta}_n$ is computed as 
\begin{equation}
\label{Theta-hat-RB}
\widehat{\Theta}_n =
\sum_{i=1}^N
w_n^i
\int
\phi({\bf x}_n) 
\underbrace{
p({\bf x}_{n}|{\bf r}_{0:n}^i,{\bf y}_{0:n})
}_
{\rm KF}
{\rm d}{\bf x}_{n},
\end{equation}
in which
$p({\bf x}_{n}|{\bf r}_{0:n}^i,{\bf y}_{0:n})$
is computed exactly via KF 
if model \eqref{hmc-jm}, conditionally on ${\bf r}_{0:n}$,
is linear and Gaussian, i.e. if
$f_{i+1|i}({\bf x}_{i+1}|{\bf x}_{i},r_{i+1})$
and 
$g_i({\bf y}_{i}|{\bf x}_{i},r_i)$ are Gaussian
with means linear in ${\bf x}_{i}$
\cite{doucet-jump-Mkv}.
\end{itemize}

\subsection{Bayesian CMC algorithms for linear and Gaussian JMSS models}
\label{linear-gaussian-jmss}

\subsubsection{Deriving the Bayesian CMC algorithm}
\label{derivation-jmss1}

In this section we begin with the second class of algorithms.
Let us first see that
\eqref{Theta-RB}
coincides with 
\eqref{def-de-Theta-jointe}
(in which the integral 
is replaced by a sum,
since $R_n$ is discrete),
up to the identification:
$X_1 = {\bf R}_{0:n-1}$,
$X_2 = R_{n}$,
$f(x_1,x_2) = \int
\phi({\bf x}_n) 
p({\bf x}_{n}|{\bf r}_{0:n-1},r_n,{\bf y}_{0:n})
{\rm d}{\bf x}_{n}$,
and
$p(x_1,x_2)$
is the 
joint pdf
\begin{equation}
\label{p_12-jump}
p({\bf r}_{0:n}|{\bf y}_{0:n}) =
\underbrace{
p({\bf r}_{0:n-1}|{\bf y}_{0:n})
}_{p(x_1)}
\underbrace{
p(r_{n}|{\bf r}_{0:n-1},{\bf y}_{0:n})
}_{p(x_2|x_1)}.
\end{equation}
We need to compute 
both factors
(we cannot simply apply the results of \S \ref{deriving-hmc},
because in 
\eqref{hmc-jm}
the marginal chain $(R_n,{\bf Y}_n)$ is not an HMC,
as was $({\bf X}_n,{\bf Y}_n)$ in  \eqref{hmc}).
We first need an approximation of $p(x_1)$, i.e. of
\begin{equation}
\label{p_n-1_n_jump}
p({\bf r}_{0:n-1}|{\bf y}_{0:n})
=
\frac
{
p({\bf y}_{n}|{\bf r}_{0:n-1},{\bf y}_{0:n-1})
p({\bf r}_{0:n-1}|{\bf y}_{0:n-1})
}
{
\sum_{{\bf r}_{0:n-1}}
p({\bf y}_{n}|{\bf r}_{0:n-1},{\bf y}_{0:n-1})
p({\bf r}_{0:n-1}|{\bf y}_{0:n-1})
}.
\end{equation}
However the SMC algorithm propagates approximations of 
$p({\bf r}_{0:n}|{\bf y}_{0:n})$.
So let 
$\widehat{p}({\bf r}_{0:n-1}|{\bf y}_{0:n-1}) =$
$\sum_{i=1}^N w_{n-1}^i \delta_{{\bf r}_{0:n-1}^i}$;
applying again Rubin's SIR mechanism,
$\widehat{p}({\bf r}_{0:n-1}|\y_{0:n}) = 
\sum_{i=1}^N
\widetilde{w}_{n-1}^i \delta_{{\bf r}_{0:n-1}^i}
$,
where
\begin{equation}
\label{w-optimal-jump}
\widetilde{w}_{n-1}^i \propto
{w}_{n-1}^i 
p(\y_n|\y_{0:n-1},{\bf r}_{0:n-1}^i) , \;\;
\sum_{i=1}^N \widetilde{w}_{n-1}^i =1,
\end{equation}
is an MC approximation of 
$p({\bf r}_{0:n-1}|\y_{0:n})$.
Next from \eqref{hmc-jm},
the second factor 
$p(x_2|x_1)$
of \eqref{p_12-jump}
can be rewritten as (here 
${\cal N}$ stands for numerator):
\begin{eqnarray}
\label{optimal-SIR-JM}
p(r_n|{\bf r}_{0:n-1}^i,\y_{0:n} ) & = &
\frac{ p(\y_n|\y_{0:n-1},{\bf r}_{0:n-1}^i,r_n)p(r_n| r_{n-1}^i)} {p(\y_n|\y_{0:n-1},\r_{0:n-1}^i) = \sum_{r_n} {\cal N}} .
\end{eqnarray}
Note that as in section \ref{semi-exact-hmc},
$p(r_n|{\bf r}_{0:n-1}^i,\y_{0:n} )$
is the optimal conditional IS distribution,
i.e. that which minimizes the conditional variance of the weights,
given 
${\bf r}_{0:n-1}^i$ and
$\y_{0:n}$.
Finally, setting $f({\bf r}_{0:n-1}^i,r_n)=$ ${\rm E}(\phi(\x_n)|\y_{0:n},{\bf r}_{0:n-1}^i,r_n)$,
the Bayesian CMC and crude estimators respectively read
\begin{eqnarray}
\label{bayesian-CMC-jmss}
\widetilde{\Theta}_n({\bf r}_{0:n-1}^{1:N}) 
& = & 
\sum_{i=1}^N 
\widetilde{w}_{n-1}^i({\bf r}_{0:n-1}^{1:N})
\sum_{r_n}
f({\bf r}_{0:n-1}^i,r_n)
p(r_n|{\bf r}_{0:n-1}^i,{\bf y}_{0:n}),
\\
\label{bayesian-crude-jmss}
\widehat{\Theta}_n({\bf r}_{0:n-1}^{1:N},{\bf r}_{n}^{1:N})
& = &
\sum_{i=1}^N 
\widetilde{w}_{n-1}^i
({\bf r}_{0:n-1}^{1:N})
f({\bf r}_{0:n-1}^i,r_n^i)
,
\end{eqnarray}
in which 
$r_n^i \sim$
$p(r_n|{\bf r}_{0:n-1}^i,{\bf y}_{0:n})$.

\subsubsection{Computing $\widetilde{\Theta}_n$ in practice: linear and Gaussian JMSS}
\label{practice-jmss2}

Implementing 
$\widetilde{\Theta}_n$
requires that 
\eqref{w-optimal-jump}
and
\eqref{optimal-SIR-JM}
are computable,
and that 
in \eqref{bayesian-CMC-jmss}
the conditional expectation 
$f({\bf r}_{0:n-1}^i,r_n)$
is computable too.
We thus need to compute
$p(\y_n|\y_{0:n-1},{\bf r}_{0:n-1}^i,r_n)$,
which is not possible in general JMSS models.
So let us now assume that the JMSS \eqref{hmc-jm}
is moreover linear 
and (conditionally) Gaussian:
\begin{eqnarray}
\label{jump1} && R_n \text{ is a discrete Markov Chain}, \\
\label{jump2} && 
{\bf X}_{n} = {\bf F}_{n}(R_{n}) {\bf X}_{n-1} + 
{\bf G}_n(R_{n}) {\bf U}_{n} , \\
\label{jump3} && 
{\bf Y}_{n} = {\bf H}_{n}(R_{n}) {\bf X}_n + {\bf L}_n (R_{n}) {\bf V}_{n} ,
\end{eqnarray}
where ${\bf X}_0$, 
${\bf U}_1, \cdots, {\bf U}_n$
and 
${\bf V}_0, \cdots, {\bf V}_n$ 
are independent and independent of $R_0, \cdots, R_n$.
We set
${\bf X}_0 \sim \mathcal{N}({\bf m}_0,{\bf P}_0)$,
${\bf U}_n \sim \mathcal{N}({\bf 0},{\bf Q}_n)$ and
${\bf V}_n \sim \mathcal{N}({\bf 0},{\bf R}^v_n)$.
Then let
$
p({\bf x}_{n-1}|{\bf r}_{0:n-1}^i, {\bf y}_{0:n-1})
=
{\cal N}
(\x_{n-1};{\bf m}_{n-1|n-1}^i;{\bf P}_{n-1|n-1}^i).
$
Then
$p({\bf y}_n|$ ${\bf y}_{0:n-1},{\bf r}_{0:n-1}^i,r_n)$
is given by the predicted observation mean and covariance of the KF,
i.e.
\begin{equation}
\label{jump-eq-cle}
p({\bf y}_n|{\bf y}_{0:n-1},{\bf r}_{0:n-1}^i,r_n)
=
\mathcal{N}({\bf y}_n;\widetilde{\bf y}_n^i(r_n);{\bf S}_n^i(r_n)),
\end{equation}
where
\begin{eqnarray}
\label{innova}
\widetilde{\bf y}_n^i(r_n)& =&{\bf y}_n-{\bf H}_n(r_n){\bf F}_n(r_n) {\bf m}_{n-1|n-1}^i \text{,} \\
\label{innovb} 
{\bf S}_n^i(r_n) & =&{\bf H}_n(r_n) {\bf P}_{n|n-1}^i(r_n) {\bf H}_n(r_n)^T + {\bf L}_n(r_n){\bf R}_n^v {\bf L}_n(r_n)^T \\
\label{cov-innov}
{\bf P}_{n|n-1}^i(r_n)&= &{\bf F}_n(r_n) {\bf P}_{n-1|n-1}^i {\bf F}_n^T(r_n) + {\bf G}_n(r_n){\bf Q}_n{\bf G}_n(r_n)^T.
\end{eqnarray}
In summary,
\eqref{jump-eq-cle}-\eqref{cov-innov}
enable to compute 
\eqref{w-optimal-jump} and
\eqref{optimal-SIR-JM}, and finally 
\eqref{bayesian-CMC-jmss}.
\begin{remark}
Estimator $\widetilde{\Theta}_n$ in 
\eqref{bayesian-CMC-jmss}
is the Bayesian CMC counterpart of $\widehat{\Theta}_n$ in 
\eqref{bayesian-crude-jmss},
which itself coincides with 
the so-called RB SMC estimator \eqref{Theta-hat-RB} 
for JMSS \cite{doucet-jump-Mkv}.
Indeed,
$\widehat{\Theta}_n$ corresponds to 
$\widehat{\Theta}^{\rm RB}$ 
in \eqref{theta-smc-2}
where $\x_1=\r_{0:n}$, $\x_2=\x_{0:n}$, $f(\x_1,\x_2)=\phi(\x_n)$ and
$q_1(r_n|\r_{0:n-1})=p(r_n|{\bf r}_{0:n-1},{\bf y}_{0:n})$.
So  \eqref{bayesian-CMC-jmss}
can be seen as a further RB step
of an already RB SMC estimator; 
the RB step 
leading to \eqref{Theta-hat-RB}
was a spatial one,
since PF was performed on variables
${\bf r}_{0:n}$,
rather than on the extended state 
$({\bf x}_{0:n}, {\bf r}_{0:n})$;
here this second RB step is temporal,
since in \eqref{bayesian-CMC-jmss}
PF acts on
${\bf r}_{0:n-1}$,
rather than on 
${\bf r}_{0:n}$.
So here is an example where we can jointly
use the classical RB-PF and our CMC Bayesian
technique; but we will see in the next section
that a CMC Bayesian estimator can also be
derived in JMSS models in which classical
RB-PF is not available.

\end{remark}

\begin{remark}
One should observe that 
if
$\widehat{\Theta}_n$ can be computed,
$\widetilde{\Theta}_n$ can be computed as well;
so the variance reduction can be achieved 
under the same assumptions (linear and Gaussian JMSS)
as those needed for the RB SMC estimator \cite{doucet-jump-Mkv}.
On the other hand 
this new variance reduction 
involves an extra computational effort,
which however 
is not prohibitive
(at least if $K$ is small),
as we see from
\eqref{bayesian-CMC-jmss} and
\eqref{bayesian-crude-jmss}.
First, weights 
$\widetilde{w}_{n-1}^i$ in
\eqref{w-optimal-jump}
have to be computed by
both algorithms.
Next for each $i$, $1 \leq i \leq N$,
both algorithms compute
$\{ p(r_n|{\bf r}_{0:n-1}^i,{\bf y}_{0:n}) \}_{r_n = 1}^{K}$.
The difference is that in the CMC algorithm
we compute directly means 
$
\sum_{r_n}
f({\bf r}_{0:n-1}^i,r_n)
p(r_n|{\bf r}_{0:n-1}^i,{\bf y}_{0:n})
$,
which requires running $K$ KF updating steps
per trajectory ${\bf r}_{0:n-1}^i$
while the crude estimator first 
extends randomly each trajectory 
before computing conditional expectations.
\end{remark}

\begin{remark}
\label{jmss-nl-soft}
Finally
our Bayesian CMC estimator  
$\widetilde{\Theta}_n$ 
stems from the RB PF which, itself,
assumes that the JMSS model 
is conditionally linear 
with additive Gaussian noise.
If this is not the case, 
but the non-linearities are not too severe, 
one can approximate 
${\rm E}(\phi(\x_n)|\y_{0:n},{\bf r}_{0:n})$ 
by EKF or UKF,
and next compute
$\widetilde{\Theta}_n$ 
from such an approximation,
by using an approximation
of $p(\y_n|\y_{0:n-1},{\bf r}_{0:n})$,
also given by EKF or UKF.
\end{remark}
\subsection{Bayesian CMC algorithms for non linear JMSS models} 
\label{extension-jmss}

In this section we derive
Bayesian CMC estimators in non linear JMSS models,
in the case where, by contrast with 
Remark \ref{jmss-nl-soft} above,
it is not possible to approximate 
${\rm E}(\phi(\x_n)|\y_{0:n},{\bf r}_{0:n})$.
In that case we need to turn back to 
the first class of SMC methods for JMSS
(see the beginning of section \ref{jmss}),
which consists in propagating
an SMC approximation of $p(\x_{0:n},\r_{0:n}|\y_{0:n})$
\cite{doucet-jump-Mkv} \cite{Andrieux-JM}.

\subsubsection{Deriving Bayesian CMC estimators} 
Let us first rewrite $\Theta_n$ as 
\begin{align}
\label{theta-non-jmss}
\Theta_n 
&= 
\sum_{\r_{0:n-1},r_{n}}\int 
\phi(\x_n) p(\x_{0:n-1},\r_{0:n-1},\x_n,r_n|\y_{0:n})  {\rm d}\x_{0:n-1} {\rm d} \x_n \text{.}
\end{align}  
Let now 
$\sum_{i=1}^N 
w_{n-1}^i
(
{\bf x}_{0:n-1}^{1:N},
{\bf r}_{0:n-1}^{1:N}
) 
\delta_{\x_{0:n-1}^i,\r_{0:n-1}^i}$
be an MC approximation of 
$p(\x_{0:n-1},$ $\r_{0:n-1}|\y_{0:n-1})$.
Then
$\widehat{p}(\x_{0:n-1}, \r_{0:n-1}|\y_{0:n})
=
\sum_{i=1}^N \widetilde{w}_{n-1}^i \delta_{\x_{0:n-1}^i,\r_{0:n-1}^i}$, 
in which
\begin{eqnarray}
\label{widetilde-1stclass}
\widetilde{w}_{n-1}^i 
& \propto &
w_{n-1}^i p(\y_n|\x_{n-1}^i,r_{n-1}^i),
\sum_{i=1}^N \widetilde{w}_{n-1}^i =1
\end{eqnarray}
is an MC approximation of $p(\x_{0:n-1}, \r_{0:n-1}|\y_{0:n})$.
Let also
\begin{equation}
\label{opt-1stclass}
(\x_n^i,r_n^i) \sim p(\x_n,r_n|\x_{n-1}^i,r_{n-1}^i,\y_n) \text{.}
\end{equation}
Then the associated crude MC estimator is
given by
\cite{doucet-jump-Mkv}
\cite{Andrieux-JM}:
\begin{eqnarray}
\label{crude-MC-JMSS} 
\widehat{\Theta}_n
(
{\bf x}_{0:n}^{1:N},
{\bf r}_{0:n}^{1:N}
) 
&=&
\sum_{i=1}^N 
\widetilde{w}_{n-1}^i
(
{\bf x}_{0:n-1}^{1:N},
{\bf r}_{0:n-1}^{1:N})
\phi(\x_n^i) ,
\end{eqnarray}

We now propose two Bayesian CMC estimators 
of $\Theta_n$,
associated to two different partitions 
of $({\bf X}_{0:n},{\bf R}_{0:n})$.
Setting
$X_1=({\bf X}_{0:n-1},{\bf R}_{0:n})$ and 
$X_2={\bf X}_n$ leads to 
$\widetilde{\Theta}_n^{{\bf X}_n}$;
setting 
$X_1=({\bf X}_{0:n-1},{\bf R}_{0:n-1})$,
$X_2=({\bf X}_n,R_n)$ and
$f(\x_{n-1}^i,\y_n,$ $\r_n)=\int  \phi(\x_n) p(\x_n|\x_{n-1}^i,r_{n},\y_n) {\rm d} \x_n$ 
leads to 
$\widetilde{\Theta}_n^{({\bf X}_n,R_n)}$,
with
\begin{eqnarray}
\label{CMC-JMSS-general-3} 
\widetilde{\Theta}^{{\bf X}_n}_n
\!
({\bf x}_{0:n-1}^{1:N},{\bf r}_{0:n}^{1:N})
& \!\!= \!\!& 
\sum_{i=1}^N 
\widetilde{w}_{n-1}^i
f(\x_{n-1}^i,\y_n,\r_n^i) ,
\\
\label{CMC-JMSS-general} 
\widetilde{\Theta}_n^{({\bf X}_n,R_n)}
({\bf x}_{0:n-1}^{1:N},{\bf r}_{0:n-1}^{1:N})
& \!\!= \!\!& 
\sum_{i=1}^N \!
\widetilde{w}_{n-1}^i \!\!
\sum_{r_n} \!\! \int \!\! \phi(\x_n) p(\x_n,r_n|\x_{n-1}^i,r_{n-1}^i,\y_n) {\rm d} \x_n
\\
& \!\!= \!\!& 
\nonumber
\sum_{i=1}^N 
\widetilde{w}_{n-1}^i\!\!
\sum_{r_n} 
p(r_n|\x_{n-1}^i,r_{n-1}^i,\y_n) 
f(\x_{n-1}^i,\y_n,\r_n),
\end{eqnarray}
in which 
$\widetilde{w}_{n-1}^i
(
{\bf x}_{0:n-1}^{1:N},
{\bf r}_{0:n-1}^{1:N}
)$
is given by \eqref{widetilde-1stclass},
and in \eqref{CMC-JMSS-general-3} 
$r_n^i \sim p(r_n|\x_{n-1}^i,$ $r_{n-1}^i,\y_n)$
(a marginal of \eqref{opt-1stclass}).
Of course,
$
\widetilde{\Theta}^{{\bf X}_n}_n
=
{\rm E}(\widehat{\Theta}_n|\x_{0:n-1}^{1:N},r_{0:n}^{1:N},\y_{0:n})$
and 
$\widetilde{\Theta}_n^{({\bf X}_n,R_n)}
=
{\rm E}(\widetilde{\Theta}^{{\bf X}_n}_n|
\x_{0:n-1}^{1:N},r_{0:n-1}^{1:N},\y_{0:n})$,
so 
${\rm var}(\widetilde{\Theta}_n^{({\bf X}_n,R_n)}) \leq 
 {\rm var}(\widetilde{\Theta}_n^{{\bf X}_n}) \leq 
 {\rm var}(\widehat{\Theta}_n)$.

\subsubsection{Computing 
$\widetilde{\Theta}_n^{{\bf X}_n}$ and
$\widetilde{\Theta}_n^{({\bf X}_n,R_n)}$
in practice}
\label{practice-jmss1}

Let us now discuss when 
\eqref{CMC-JMSS-general-3} and
\eqref{CMC-JMSS-general} can be computed. 
In model \eqref{hmc-jm},
$
p(\y_n|\x_{n-1}^i,r_{n-1}^i)= \sum_{r_n} p(r_n|r_{n-1}^i)p(\y_n|\x_{n-1}^i,r_n)
$
and
$p(\y_n|\x_{n-1}^i,r_{n-1}^i) p(\x_n,$ $r_n|\x_{n-1}^i,r_{n-1}^i,\y_n)=
p(r_n|r_{n-1}^i)p(\y_n|\x_{n-1}^i,r_n)p(\x_n|\x_{n-1}^i,r_n,\y_n)$.
So let
\begin{eqnarray}
\label{poids-CMC-JMSS}
\overline{w}_{n-1}^i(r_n) 
& = & 
\frac
{w_{n-1}^i p(r_n|r_{n-1}^i) p(\y_n|\x_{n-1}^i,r_n)}
{\sum_{r_n}\sum_{i=1}^N 
 w_{n-1}^i p(r_n|r_{n-1}^i) p(\y_n|\x_{n-1}^i,r_n^i)}
\end{eqnarray}
(note that 
$\sum_{r_n} \overline{w}_{n-1}^i(r_n) =$
$\widetilde{w}_{n-1}^i$).
Then
$\widetilde{\Theta}_n^{{\bf X}_n}$ and
$\widetilde{\Theta}_n^{({\bf X}_n,R_n)}$
can be rewritten as 
\begin{eqnarray}
\label{CMC-JMSS-general-32}
\widetilde{\Theta}_n^{{\bf X}_n}
& = & 
\sum_{i=1}^N 
\left[ \sum_{r_n} \overline{w}_{n-1}^i(r_n) \right] 
\int \phi(\x_n) p(\x_n|\x_{n-1}^i,\y_n,r_n^i) {\rm d} \x_n ,
\\
\label{CMC-JMSS-general-2}
\widetilde{\Theta}_n^{({\bf X}_n,R_n)}
& = & 
\sum_{i=1}^N 
\left[\sum_{r_n}
\overline{w}_{n-1}^i(r_n) 
\int \phi(\x_n) p(\x_n|\x_{n-1}^i,\y_n,r_n) {\rm d} \x_n \right].
\end{eqnarray}
So \eqref{CMC-JMSS-general-2} is computable
as soon as \eqref{CMC-JMSS-general-32} is computable.
On the other hand,
$\widetilde{\Theta}_n^{{\bf X}_n}$
is a generalization of \eqref{bayesian-CMC-sir}:
$p(\x_n|\x_{n-1},\y_n,r_n)$ and $p(\y_n|\x_{n-1},r_n)$
play the same role as 
$p(\x_n|\x_{n-1},\y_n)$ and $p(\y_n|\x_{n-1})$ 
in \S 
\ref{deriving-hmc}
except that we have now introduced a dependency in $r_n$. 
This means that 
$\widetilde{\Theta}_n^{{\bf X}_n}$ and
$\widetilde{\Theta}_n^{({\bf X}_n,R_n)}$ 
are computable 
as soon as the Bayesian CMC estimator 
\eqref{bayesian-CMC-sir}
of \S \ref{deriving-hmc}
is computable in the underlying HMC model
(i.e., the HMC model to which the JMSS reduces  when the jumps are known),
see section \ref{exact-computation-CMC-hmc}.
For example, semi-linear stochastic models (including the ARCH ones)
with Markov jumps are a class of models
in which 
\eqref{CMC-JMSS-general-32} and
\eqref{CMC-JMSS-general-2} 
are computable.

Finally the only difference between 
\eqref{CMC-JMSS-general-32} and
\eqref{CMC-JMSS-general-2}
comes from the computational cost
that we discuss now. 
For a given $i$,
in \eqref{CMC-JMSS-general-2}
the computation of $\int \phi(\x_n) p(\x_n|\x_{n-1}^i,r_n,\y_n) {\rm d}\x_n$
has to be done for all $r_n$,
while in \eqref{CMC-JMSS-general-32}
it has only to be done
for the $r_n^i$ which has been sampled.
So as expected, 
$\widetilde{\Theta}_n^{({\bf X}_n,R_n)}$ is preferable to 
$\widetilde{\Theta}_n^{{\bf X}_n}$ 
but requires an extra computational cost.
On the other hand,
comparing the computational cost of
$\widetilde{\Theta}_n^{{\bf X}_n}$ and 
$\widehat{\Theta}_n$
is the same issue as comparing
that of the Bayesian CMC estimator \eqref{bayesian-CMC-sir}
to that of the crude MC one \eqref{bayesian-crude-sir}
in Section \ref{computing-practice},
and is thus problem dependent. 
However, one should observe that 
in the particular case described at the end of 
section \ref{exact-computation-CMC-hmc}, 
i.e. 
when sampling according to $p(\x_n|\x_{n-1},r_n,\y_n)$
requires the computation of 
$\int \phi(\x_n) p(\x_n|\x_{n-1},r_n,\y_n) {\rm d}\x_n$,
then the computation of
$\widetilde{\Theta}_n^{{\bf X}_n}$ 
does not involve an extra computational cost 
as compared to that of $\widehat{\Theta}_n$.

\subsubsection{Approximate computation}
\label{approximate-class1}

Let us finally discuss on approximate computation of
$\widetilde{\Theta}_n$ 
when 
$p(\y_n|\x_{n-1},r_n)$ and
$p(\x_n|\x_{n-1},\y_n,r_n)$
in \eqref{CMC-JMSS-general-2} are not available.
First notice that 
\eqref{CMC-JMSS-general-2} can be computed
with
the same numerical approximations
as those which were used 
in the computation of \eqref{bayesian-CMC-sir}
(see 
section \ref{approximate-computation-CMC-hmc} above),
except that they have
to be done for all possible values of $r_n$. 
However, $r_n$ is discrete and as we now see,
one can derive other approximation techniques:
\begin{itemize}
\item
In \eqref{hmc-jm} we have 
$p(r_n,\x_n,\y_n|\x_{n-1},r_{n-1})=p(r_n|r_{n-1})f_{n|n-1}(\x_n|\x_{n-1},r_n)$ 
$g_n(\y_n|\x_n,r_n)$
so the numerator of \eqref{CMC-JMSS-general-2}
can be rewritten as
$$
\sum_{r_n} \int \phi(\x_n) \left[ \sum_{i=1}^N p(r_n|r_{n-1}^i) w_{n-1}^i f_{n|n-1}(\x_n|\x_{n-1}^i,r_n)
g_n(\y_n|\x_{n},r_n) \right] {\rm d}\x_n.
$$
If for a given 
$r_n$ the integral is not computable,
one can approximate it with IS
by
sampling
$\x_n^{r_n,i} \sim q(\x_n|\x_{n-1}^i,r_n)$
for all $i$, $1 \leq i \leq N$
and for all $r_n$.
An approximation of the numerator is then given by 
$\sum_{r_n}  \sum_{i=1}^N \phi(\x_n^{r_n,i})  $ $ p(r_n|r_{n-1}^i) w_{n-1}^i f_{n|n-1}(\x_n^{r_n,i}|\x_{n-1}^i,r_n)
g_n(\y_n|\x_{n}^{r_n,i},r_n)/q(\x_n^{r_n,i}|\x_{n-1}^i,r_n)$. So
we do not use samples for the discrete part $r_n$.
We apply the same approximation
for the denominator which can be rewritten as
$\sum_{r_n}\sum_{i=1}^N w_{n-1}^i$ $ p(r_n|r_{n-1}^i)
\times  \int f_{n|n-1}(\x_n|r_n,\x_{n-1}^i) g_n(\y_n|\x_{n},r_n) {\rm d} \x_n$;   

\item
When the optimal distribution  $p(\x_n,r_n|\x_{n-1},r_{n-1},\y_n)$ 
is not available, it has been proposed \cite{doucet-jump-Mkv} to
sample independently $(\x_n^i,r_n^i)$
according to an importance distribution 
$q(\x_n,r_n|\x_{n-1}^i,r_{n-1}^i)=q(r_n|\x_{n-1}^i,r_{n-1}^i)q(\x_n|\x_{n-1}^i,$
$r_n,r_{n-1}^i)$,
for all $i$, $1 \leq i \leq N$,
then to compute the estimator
$\widehat{\Theta}^{\rm SIR}_n
=
\sum_{i=1}^N \frac{w_n^i}{\sum_{i=1}^N w_n^i} \phi(\x_n^i)
$,
$w_n^i
=
w_{n-1}^i \frac{ p(r_n^i|r_{n-1}^i)f_{n|n-1}(\x_n^i|r_{n}^i,\x_{n-1}^i)g_n(\y_n|\x_n^i,r_n^i)}{q(\x_n^i,r_n^i|\x_{n-1}^i,r_{n-1}^i)}
$.
Remember that the Bayesian CMC estimator $\widetilde{\Theta}_n$
is actually the expectation of the crude MC estimator
$\widehat{\Theta}_n$
given some variables. 
One can wonder if it is not possible to compute  
${\rm E}(\widehat{\Theta}_n^{\rm SIR}|\{r_{n-1}^i,\x_{n-1}^i,\x_n^i\}_{i=1}^N,\y_n)$,
i.e. to compute the expectation of 
$\widehat{\Theta}^{\rm SIR}_n$
as a function of $r_n$. 
Since $\phi(.)$ does not depend on $r_n$, 
it is equivalent
to compute the conditional
expectation of the unnormalized weights
$w_n^i$
and so to reduce their variance. 
Unfortunately this is not possible because
of the normalization factor. 
However one can 
compute separately the conditional expectation of the numerator 
and that of the denominator.
This 
is an easy task since $r_n$ takes its values in a discrete set,
and $q(r_n|\x_n,\x_{n-1},r_{n-1})\! \! = \! q(r_n|\x_{n-1},r_{n-1})q(\x_n|\x_{n-1},r_n,r_{n-1}) \! / \! \sum_{r_n}\! q(r_n|
\x_{n-1},$ $r_{n-1})q(\x_n|\x_{n-1},r_n,r_{n-1})$.
This variance reduction of the unnormalized weights
comes from a normalized IS implementation of
\eqref{CMC-JMSS-general-2} which is rewritten as 
\begin{align}
& \widetilde{\Theta}_n =
\nonumber \\
& 
\frac{
\int \phi(\x_n)  \sum_{i=1}^N \left[\sum_{r_n}p(r_n|r_{n-1}^i) w_{n-1}^i f_{n|n-1}(\x_n|\x_{n-1}^i,r_n)
g_n(\y_n|\x_{n},r_n) \right] {\rm d}\x_n}
{\int \sum_{i=1}^N \left[\sum_{r_n}p(r_n|r_{n-1}^i) w_{n-1}^i f_{n|n-1}(\x_n|\x_{n-1}^i,r_n)
g_n(\y_n|\x_{n},r_n) \right] {\rm d}\x_n} \text{}
\end{align}
with the importance distribution
$$q(\x_n|\x_{n-1},r_{n-1})=\sum_{r_n} q(r_n|\x_{n-1},r_{n-1})q(\x_n|\x_{n-1},r_n,r_{n-1}) \text{.}$$
Note that the computation of the new weights is not prohibitive 
as long as $K << N$.

\end{itemize}

\subsection{Simulations}
\label{simu-jmss}

We now test our approach 
in a linear and Gaussian JMSS model,
described by equations \eqref{jump1}-\eqref{jump3}
in which 
$${\bf F}_n(r)=\begin{bmatrix}
1 & \frac{ \sin(\omega_r T)}{\omega_r} & 0 & - \frac{1-\cos(\omega_r T)}{\omega_r} \\
0 & \cos(\omega_r T) & 0 & - \sin(\omega_r T) \\
0 & \frac{1-\cos(\omega_r T)}{\omega_r} & 1 & \frac{ \sin(\omega_r T)}{\omega_r} \\
0 &  \sin(\omega_r T) & 0 & \cos(\omega_r T)
\end{bmatrix} \text{, }
{\bf Q}_n= \sigma_v^2 \begin{bmatrix}
\frac{T^3}{3}&\frac{T^2}{2}&0&0 \\
\frac{T^2}{2}&T&0&0 \\
0&0&\frac{T^3}{3}&\frac{T^2}{2} \\
0&0&\frac{T^2}{2}&T 
\end{bmatrix} \text{, }$$
${\bf H}_n=\begin{bmatrix}
1 & 0 & 0 & 0 \\
0 & 0 & 1 & 0 
\end{bmatrix}$, 
${\bf R}_n=\begin{pmatrix}
\sigma_x ^2 & 0 \\
0 & \sigma_y ^2 
\end{pmatrix} \text{ and}$
${\bf G}_n(r)= {\bf I}$,
$T=2s$, $\sigma_v=3m^2/sec^3$
and $\sigma_x= \sigma_y=10m$.
We track a maneuvering target
described by its position and velocity in the Cartesian 
coordinates, 
$\x_n=[p_x,\dot{p}_x,p_y,\dot{p}_y]^T_n$.
Mode $r_n$ represents the behavior of the target:
straight, left turn and right turn.
Remember from \ref{practice-jmss2} that the computation
of $\widetilde{\Theta}_n$ involves an extra computational 
cost compared to that of $\widehat{\Theta}_n$. So 
we compute the efficiency over $P=200$ simulations
defined as \cite{lecuyer}
\begin{align}
{\rm Eff}(n)= \frac{1}{{\rm MSE}(n){\rm E}(C(n))} \text{,}
\end{align}
where $C(n)$ is the CPU
time to compute the estimator,
and we discuss the performances of
$\widehat{\Theta}_n$ and
$\widetilde{\Theta}_n$ 
in function of the model parameters. 
Both  estimators are computed
with $N=1000$ particles. 

\begin{figure} [htbp]
\center
\epsfig{file=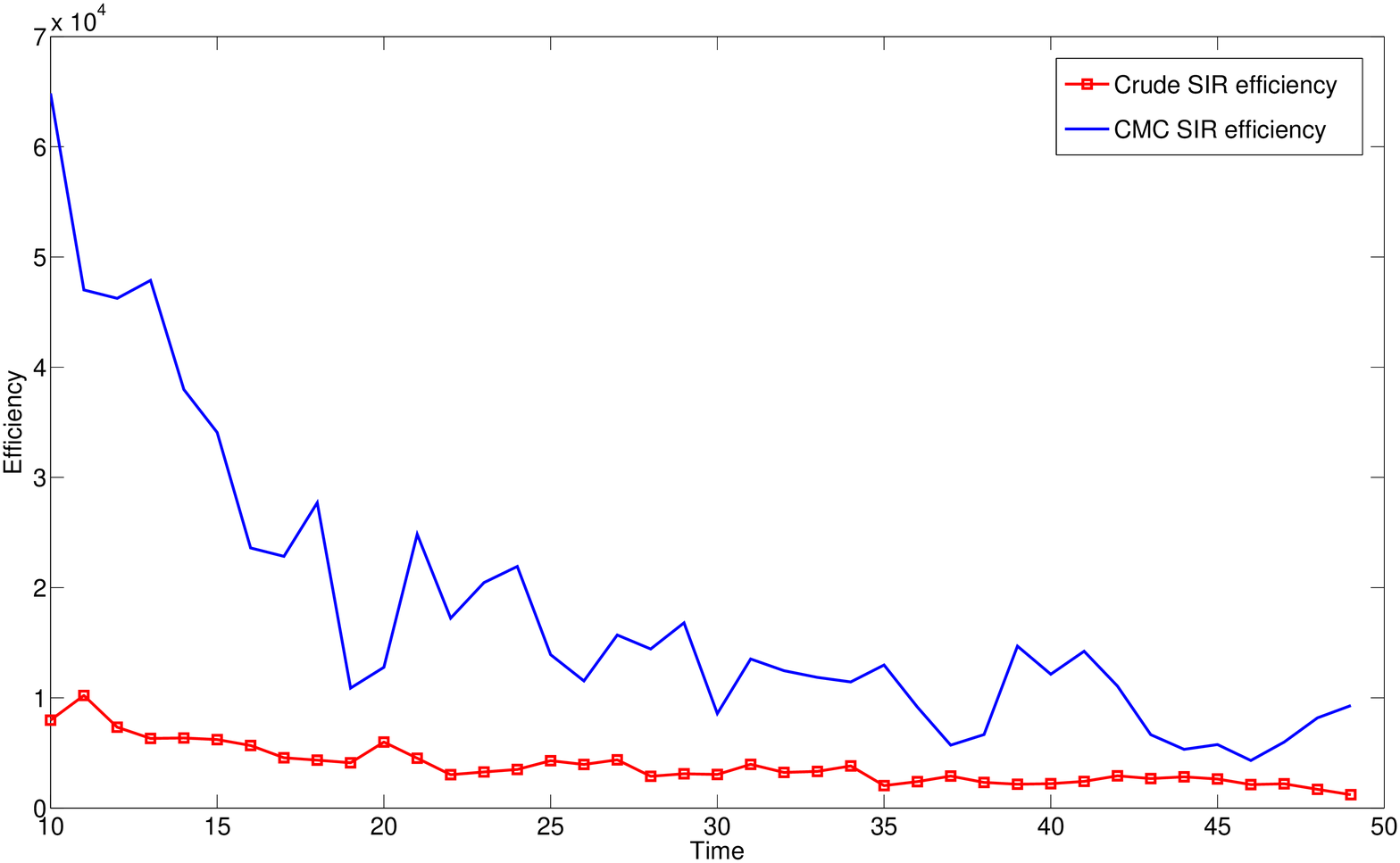,height=0.2\textheight,width=0.85\textwidth}
	\caption{Efficiency - Linear JMSS Model - Close Markovian transition probabilities - $\phi(\x_n)=\x_n$. Due to the weak dependency 
	of the model in $\r_n$, it is dangerous to sample new particles (Crude SIR) before computing the estimator. }
	\label{efficiency1}
\end{figure}

We first set 
$\omega_1=0 \text{ rad}.\text{s}^{-1}$,
$\omega_2= 3 \pi / 180  \text{ rad}.\text{s}^{-1} $ and
$\omega_3= - 3 \pi / 180  \text{ rad}.\text{s}^{-1}$.
The Markovian transition probability is 
$p(r_n|r_{n-1})=0.4$ if $r_n=r_{n-1}$ and
$p(r_n|r_{n-1})=0.3$ otherwise.
In Figure \ref{efficiency1}, we display the
(averaged) efficiency of both estimators over time.  
The efficiency of $\widetilde{\Theta}_n$ is greater
than that of $\widehat{\Theta}_n$, 
so the Bayesian CMC estimator for linear JMSS is of practical interest.
Note that the dependency of the model 
in $\{r_n\}$ is weak since $w_r$ is small and
the Markovian transition probabilities are close.
So distributions $\{p(r_n|\r_{0:n-1}^i,\y_{0:n})\}_{i=1}^N$
tend to be uniform, and remember that $\widetilde{\Theta}_n$
computes directly the expectations
according to these distributions
while $\widehat{\Theta}_n$ uses samples 
${\bf r}_n^{1:N}$ according to them.
This is why the gap between both estimators gets larger 
when distributions $\{p(r_n|\r_{0:n-1}^i,\y_{0:n})\}_{i=1}^N$
become almost uniform.

\begin{figure} [htbp]
\center
\epsfig{file=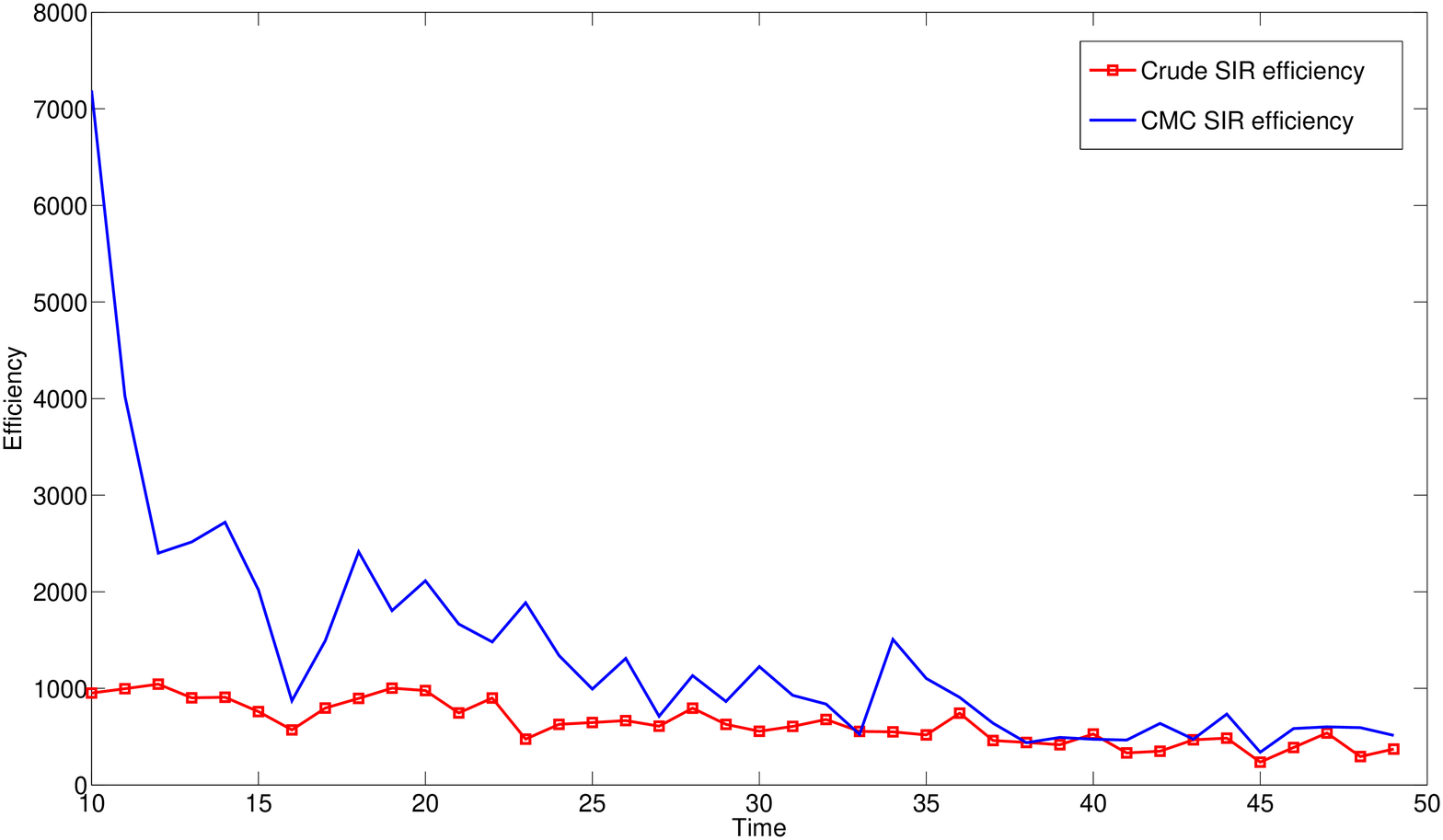,height=0.2\textheight,width=0.85\textwidth}
	\caption{Efficiency - Linear JMSS Model -  Dispersed Markovian transition probabilities - $\phi(\x_n)=\x_n$. Contrary
	to the previous simulation, distributions $\{p(r_n|\r_{0:n-1}^i,\y_{0:n})\}_{i=1}^N$ are dispersed so the gap between the crude SIR estimator
	and the CMC SIR one shrinks. }
	\label{efficiency2}
\end{figure}

Next we increase the dependency of the model in $\{r_n\}$
by setting 
$\omega_1= 0  \text{ rad}.\text{s}^{-1} $,
$\omega_2= 8 \pi / 180  \text{ rad}.\text{s}^{-1} $
and  $\omega_3= - 8 \pi / 180  \text{ rad}.\text{s}^{-1}$. 
We also set $p(r_k|r_{k-1})=0.6$ if $r_k=r_{k-1}$ and
$p(r_k|r_{k-1})=0.2$ otherwise.
In Figure \ref{efficiency2}, we display 
the averaged efficiency of both estimators
for this new set of parameters. Indeed,
the gap between both estimators is reduced
but $\widetilde{\Theta}_n$ still outperforms
$\widehat{\Theta}_n$.

\section{Bayesian CMC algorithms for Multi-Target filtering}
\label{multi-phd}
In this final section we apply CMC to multi-target filtering.
Some adaptations are necessary, because in the multi-target
context we do not necessary deal with classical pdf.
However, the discussion in section \ref{bayesian-cmc-estimators}
still holds, as we shall see. 
Let us begin with a brief review
of multi-object filtering.

\subsection{A brief review of Random Finite Sets (RFS) based multi-target filtering}
\label{review-RFS}
Multi-object filtering extends the previous problem
in the sense that we now look for estimating an unknown
number of targets from a set of observations
which are either due to detected targets
or are false alarms measurements.
Classical solutions such as
the Joint Probabilist Data association filter \cite{SHALOM_JPDAshort}
or the Multiple Hypothesis Tracker \cite{BLACKMAN}
include a matching mechanism
between targets and observations.
Alternate solutions are based on RFS,
which are sets of random variables
with random and time-varying
cardinal
(see e.g. \cite{MAHLER_LIVRE2007}).
The interest of RFS based techniques over classical solutions
is that they no longer require such a matching mechanism.
The RFS formulation was first used
to derive the multi-object Bayesian filter,
which generalizes the classical
single object one \cite{MAHLER_ARTICLE2003}.
This multi-object Bayesian filter
involves the computation
of set integrals of multi-object densities,
i.e. of positive functions $f(X)$ of a given RFS $X$,
and cannot be computed in practice
(SMC approximations can however be of interest
when the number of targets is small
\cite{VO_SMC_PHD}).
Later on,
Mahler proposed to propagate
a first order moment of the multi-object density,
the so-called PHD or intensity
\cite{MAHLER_ARTICLE2003}.
Let $|X \cap S|$ be the number of objects in RFS $X$
which belong to region $S$;
then the PHD $v(\x)$ is
defined as
the spatial density
of the expected number of targets, i.e.
\begin{align}
\label{def-phd}
\int_{S \ \subset \RR^p} v(\x)\, \mathrm {\rm d} \x =  {\rm E}(|X \cap S| )\text{.}
\end{align}
Its interest in multi-object filtering is twofold;
first,
$ \int v(\x) {\rm d} \x$
is an estimate
of the number of targets;
in addition, extracting the states
consists in looking for regions
where the PHD is high, and so local
maxima of $v$ are required.
Let now $v_{n}(\x)$
be the a posteriori PHD,
i.e. the first order moment $v_{n}(\x)$
of the multi-object density at time $n$,
given the set of past measurements
$Z_{0:n}=\{Z_1,\cdots,Z_n\}$,
where
$Z_k$ is the set of measurements available at time $k$.
The PHD filter is a set of equations which enables to propagate $v_n$
and which has the advantage
to make use of classical integrals only.
If we assume that the cardinality distributions
of the number of targets and of false alarm measurements are Poisson,
and that each target evolves and generates observations
independently of one another,
then PHD $v_n$ is propagated as follows
(we assume for simplicity
that there is no spawning)
\cite{MAHLER_ARTICLE2003}
\cite{MAHLER_LIVRE2007}:
\begin{eqnarray}
\label{phd-predic}
v_{n|n-1}(\x_n)  & = & \!\! \int \!\!  p_{s,n}(\x_{n-1}) f_{n|n-1}(\x_n|\x_{n-1}) v_{n-1}(\x_{n-1})  {\rm d}\x_{n-1} 
+ \gamma_n(\x_n) \text{,} \\
\nonumber 
v_{n}(\x_n)    & = & \left[1-p_{d,n}(\x_n)\right]  v_{n|n-1}(\x_n) \\
\label{phd-update}
& + &
\sum_{\z\in Z_n} \frac { p_{d,n}(\x_n)g_n(\z|\x_n)v_{n|n-1}(\x_n) } {\kappa_n(\z)  
+  
\int p_{d,n}(\x_n)g_n(\z|\x_n)v_{n|n-1}(\x_n) \,  {\rm d}\x_n} \text{,}
\end{eqnarray} 
where
$p_{s,n}(.)$ (resp. 
$p_{d,n}(.)$)
is the probability 
of survival (resp. 
of detection)
at time $n$ 
which can depend on state $\x_{n-1}$
(resp. on $\x_{n}$); 
and
$\kappa_n(.)$ (resp.
$\gamma_n(.)$)
is the intensity of the false alarms measurements (resp. of the birth targets) at time $n$.

\subsection{Deriving the Bayesian CMC PHD estimator}
\label{CMC-PHD}

Let us now turn back to the derivation
of a Bayesian CMC PHD estimator.
First, the problem we address 
is to compute the moment
$\Theta_n=\int f(\x_n)v_{n}(\x_n) {\rm d}\x_n$
(typically, we shall take either $f(\x_n)=1$ 
or $f(\x_n)={\bf 1}_S(\x_n)$, where $S$ is some region of interest).
From now on we assume that $p_{d,n}$ 
does not depend on $\x_n$.
Plugging \eqref{phd-predic} in 
\eqref{phd-update}, 
the PHD at time $n$ can be written as
\begin{equation}
\label{decomposition-somme-PHD}
v_{n}(\x_n) 
=
\sum_{i=1}^4 v_{n}^i(\x_n) \text{,}
\end{equation}
where
\begin{align}
v_{n}^1(\x_n)&= \left[1-p_{d,n}\right] \int p_{s,n}(\x_{n-1}) f_{n|n-1}(\x_n|\x_{n-1}) v_{n-1}(\x_{n-1})  {\rm d}\x_{n-1} \text{,} \\
v_{n}^2(\x_n)&= \left[1-p_{d,n}\right] \gamma_n(\x_n) \text{,} \\
v_{n}^3(\x_n) \! \! &= \!\!\! \! \! \sum_{\z\in Z_n}  \!\! \frac { \! p_{d,n}g_n(\z|\x_n) \!\! \int \! p_{s,n}(\x_{n-1}) \! f_{n|n-1}(\x_n|\x_{n-1})  v_{n-1}(\x_{n-1})  {\rm d}\x_{n-1} } {B_n(\z)} \text{,} \\
v_{n}^4(\x_n)&= \sum_{\z\in Z_n} \frac { p_{d,n}g_n(\z|\x_n) \gamma_n(\x_n)}{B_n(\z)} \text{,}
\end{align}
and where
\begin{align}
B_n(\z)&=  \kappa_n(\z) + B^1_n(\z)+ B^2_n(\z)  \text{,} \\
B^1_n(\z) \! &= \! \int p_{d,n}g_n(\z|\x_n) \int p_{s,n}(\x_{n-1}) f_{n|n-1}(\x_n|\x_{n-1}) v_{n-1}(\x_{n-1}) {\rm d} \x_{n-1} {\rm d}\x_n \nonumber  \\
&= p_{d,n} \int p_{s,n}(\x_{n-1})p(\z|\x_{n-1}) v_{n-1}(\x_{n-1}) {\rm d} \x_{n-1} \text{,} \\
B^2_n(\z)&= \int p_{d,n}g_n(\z|\x_n) \gamma_n(\x_n) {\rm d}\x_n \text{.}
\end{align}
Term $v_{n}^1$ (resp. $v_{n}^2$) is due to non-detected persistent (resp. birth)
targets, while $v_{n}^3$ (resp. $v_{n}^4$) is due 
to detected persistent (resp. birth) targets.

From 
\eqref{decomposition-somme-PHD} we see that
\begin{equation}
\label{decomposition-somme-Theta-PHD}
\Theta_n=
\sum_{i=1}^4
\underbrace{\int f(\x_n) v_{n}^i(\x_n) {\rm d}\x_n}
_{\Theta_n^i}
,
\end{equation}
so we now consider 
whether one can adapt the Bayesian CMC methodology 
of section 
\ref{bayesian-cmc-estimators}
to any of the moments $\Theta_n^i$.
First,
note that
$v_{n}^2(\x_n)$ and $v_{n}^4(\x_n)$ 
do not depend on $v_{n-1}(\x_{n-1})$ 
so we use a crude MC procedure
to compute $\Theta_{n}^2$ and $\Theta_{n}^4$. 
Let 
$\widehat{v}_{n-1}
= \sum_{i=1}^{L_{n-1}} w_{n-1}^i \delta_{\x_{n-1}^i} 
$ 
and 
$\widehat{\gamma}_{n} 
= \sum_{i=1}^{L_{\gamma_n}} w_{\gamma_n}^i
\delta_{\x_{\gamma_n}^i} 
$
be MC approximations of 
$v_{n-1}(\x_{n-1})$ and of $\gamma_n(\x_n)$,
respectively.
Then 
$ \widehat{\Theta}_{n}^2
= 
\sum_{i=1}^{L_{\gamma_n}} w_{n}^{2,i} f(\x_{\gamma_n}^i) \text{,}$
$  
\widehat{\Theta}_{n}^4
=
\sum_{i=1}^{L_{\gamma_n}} \left[ \sum_{\z\in Z_n} w_{n}^{4,i}(\z)\right] f(\x_{\gamma_n}^i) \text{,}
$
where
$
w_{n}^{2,i}= w_{\gamma_n}^i \left [1-p_{d,n} \right]   \text{,} 
$
$
w_{n}^{4,i}(\z)= w_{\gamma_n}^i   \frac{ p_{d,n} g_n(\z|\x_{\gamma_n}^i)}{\widetilde{B}_n(\z)}  \text{ and } \\
$
\begin{align}
\label{tilde_B}
\widetilde{B}_n(\z)\!\! &= \! \!\kappa_n(\z) \! +  \! p_{d,n}  \!\! \sum_{i=1}^{L_{\gamma_n}} \! w_{\gamma_n}^i g_n(\x_{\gamma_n}^i|\z)  \! + \! p_{d,n} \! \sum_{i=1}^{L_{n-1}}\!\! \! w_{n-1}^i p_{s,n}(\x_{n-1}^i) p(\z|\x_{n-1}^i) \text{.} 
\end{align}

By contrast, 
the computation of $v_{n}^1(\x_n)$
and of $v_{n}^3(\x_n)$
depends on $v_{n-1}(\x_{n-1})$. 
This suggests adapting the common methodology
described in section \ref{sec:intro}, 
even though the PHD is not a pdf
(it is a positive function,
but remember from \eqref{def-phd} that its integral is not equal to $1$),
and that weights $\{w_{n-1}^i\}_{i=1}^{L_{n-1}}$ may depend on variables
different from $\x_{n-1}^{1:L_{n-1}}$,
but which are known at time $n-1$. 
These differences do not impact the 
discussion of section \ref{bayesian-cmc-estimators}
which can be used in this context.
Indeed, 
we have
$f_{n|n-1}(\x_n) g_n(\z|\x_n) = 
p(\x_n|\x_{n-1},\z) p(\z|\x_{n-1})$,
so $\Theta_{n}^1$ and $\Theta_{n}^3$
can be rewritten as
\begin{eqnarray}
\label{PHD-Thetan1}
\Theta_{n}^1 \!\!
&=& \!\! 
\left[1-p_{d,n} \right ] \int {\rm E}(f(\x_n)|\x_{n-1}) \times \left[ p_{s,n}(\x_{n-1})  v_{n-1}(\x_{n-1}) \right ]{\rm d}\x_{n-1} ,
\\
\label{PHD-Thetan3}
\Theta_{n}^3 \!\!
&=& \!\! \!
\sum_{\z \in Z_n} \! \!\! \int  \!\! {\rm E}(f(\x_n)|\x_{n-1},\z) \!\left[ \frac{ p_{d,n} p_{s,n}(\x_{n-1}) p(\z|\x_{n-1})v_{n-1}(\x_{n-1})}{B_n(\z)} \right ] \!\! {\rm d}\x_{n-1} .
\end{eqnarray}

Let us start with the computation 
of $\Theta_{n}^1$ in \eqref{PHD-Thetan1}. 
Even if is not a pdf,
the factor 
$p_{s,n}(\x_{n-1})  v_{n-1}(\x_{n-1})$
within brackets
plays the role of $p(x_1)$ in 
\eqref{def-de-Theta-conditionnel},
and 
can be approximated by 
$\sum_{i=1}^{L_{n-1}} w_{n}^{1,i} \delta_{\x_{n-1}^i}$
where  $w_{n}^{1,i}=\left [1-p_{d,n} \right] p_{s,n}(\x_{n-1}^i)w_{n-1}^i $. 
So the crude MC and Bayesian CMC estimators 
of $\Theta_{n}^1$ are respectively
\begin{align}
\widehat{\Theta}_{1,n}&= \sum_{i=1}^{L_{n-1}} w_{n}^{1,i} f(\x_n^{i}) \text{,} \\
\widetilde{\Theta}_{1,n}&=\sum_{i=1}^{L_{n-1}} w_{n}^{1,i} {\rm E}(f(\x_n)|\x_{n-1}^i) \text{,}
\end{align}
in which 
$\x_n^{i} \sim f_{n|n-1}(\x_n|\x_{n-1}^i)$.
Let us next address $\Theta_{n}^3$ in \eqref{PHD-Thetan3}.
For each measurement $\z \in Z_n$,
the factor 
$\frac{ p_{d,n} p_{s,n}(\x_{n-1}) p(\z|\x_{n-1})v_{n-1}(\x_{n-1})}{B_n(\z)}$
within brackets
plays the role of $p(x_1)$ in 
\eqref{def-de-Theta-conditionnel},
and 
can be approximated by
$\sum_{i=1}^{L_{n-1}} w_{n}^{3,i}(\z) \delta_{\x_{n-1}^i}$ 
where  
$w_{n}^{3,i}(\z)= 
p_{d,n} p_{s,n}(\x_{n-1}^i) p(\z|\x_{n-1}^i)w_{n-1}^{i} / 
\widetilde{B}_n(\z)$.
So the crude MC and Bayesian CMC estimators 
of $\Theta_{n}^3$ are respectively
\begin{align}
\widehat{\Theta}_{n}^3
&=
\sum_{\z} \sum_{i=1}^{L_{n-1}} w_{n}^{3,i}(\z)f(\x_n^{\z,i}) \text{,} \\
\widetilde{\Theta}_{n}^3
&= 
\sum_{\z \in Z_n} \sum_{i=1}^{L_{n-1}} w_{n}^{3,i}(\z)
{\rm E}(f(\x_n)|\x_{n-1}^i,\z) \text{,}  
\end{align}
in which
$\x_n^{\z,i} \sim p(\x_n|\x_{n-1}^i,\z)$.

In summary,
the crude MC PHD estimator $\widehat{\Theta}_n$ of $\Theta_n$ 
is the sum of four crude MC estimators:
$\widehat{\Theta}_n
=
\sum_{i=1}^4 \widehat{\Theta}_{n}^i
$,
while our Bayesian CMC PHD estimator 
$\widetilde{\Theta}_n$
is a sum of two crude MC and two Bayesian CMC estimators:
$
\widetilde{\Theta}_n
=
\widetilde{\Theta}_{n}^1 + \widehat{\Theta}_{n}^2 +
\widetilde{\Theta}_{n}^3 + \widehat{\Theta}_{n}^4
$.
Since 
$\widetilde{\Theta}_{n}^1$ and 
$\widetilde{\Theta}_{n}^3$ are computed from 
the same MC approximation of 
$v_{n-1}(\x_{n-1})$, 
$\widetilde{\Theta}_n=
{\rm E}(\widehat{\Theta}_n| \{\x_{n-1}^i\}_{i=1}^{L_{n-1}},\{\x_{\gamma_n}^i\}_{i=1}^{L_{\gamma_n}},Z_n)$,
so section \ref{sec:intro} enables to conclude
that 
$\widetilde{\Theta}_n$
indeed outperforms
$\widehat{\Theta}_n$.

\begin{remark}
The computation of 
$\widehat{\Theta}_{n}^1 
+\widehat{\Theta}_{n}^3$
involves to sample $L_{n-1}(|Z_n|+1)$ particles
where $|Z_n|$ is the cardinal of $Z_n$. 
It is possible to compute an approximation with $L_{n-1}$ 
particles by sampling 
$\x_n^i \sim q^i(x)$
with $q^i(x) \propto w_n^{1,i}f_{n|n-1}(\x_n|\x_{n-1}^i)+ \sum_{\z}w_n^{3,i}(\z)p(\x_n|\x_{n-1}^i,\z)$
for all $i$, $1 \leq i \leq L_{n-1}$,
and by taking
\begin{align}
\widehat{\Theta}_{n}^1 + 
\widehat{\Theta}_{n}^3
&= \sum_{i=1}^{L_{n-1}} \left[ w_n^{1,i} + \sum_{\z}w_n^{3,i}(\z)\right]f(\x_n^i) \text{.}
\end{align}
\end{remark}

\begin{remark}
Depending on the form of $\gamma_n(\x_n)$ and $g_n(\z|\x_n)$, 
$\int g_n(\z|\x_n) \gamma_n(\x_n) {\rm d}\x_n$ 
may be directly computable, 
so $B^2_n(\z)$, $\Theta_{2,n}$ and $\Theta_{4,n}$ 
may be computable too.
In this case one can replace $\widetilde{B}_n(\z)$ in \eqref{tilde_B} by
$\kappa_n(\z) +B^2_n(\z)  + p_{d,n} \sum_{i=1}^{L_{n-1}} w_{n-1}^i $ $p_{s,n}(\x_{n-1}^i) p(\z|\x_{n-1}^i) \text{.} $
\end{remark}

\subsection{Computing the CMC PHD filter $\widetilde{\Theta}_n$ in practice}

In the multi-target filter problem, we look for computing
an estimator of the number of targets and of multi-target states. 
From \eqref{def-phd}, an estimator of the number
of targets is given by \begin{align}
\widetilde{N}_n&= \sum_{i=1}^{L_{n-1}} w_{n}^{1,i} +  \sum_{\z \in Z_n} \sum_{i=1}^{L_{n-1}} w_{n}^{3,i}(\z)
+ \sum_{i=1}^{L_{\gamma_n}} w_{n}^{2,i}
+  \sum_{\z\in Z_n} \sum_{i=1}^{L_{\gamma_n}} w_{n}^{4,i}(\z).
\end{align}
The procedure to extract persistent targets 
consists in looking for local maxima of 
$\sum_{i=1}^{L_{n-1}} w_{n}^{1,i} p(\x_n|\x_{n-1}^i) + 
\sum_{\z \in Z_n} \sum_{i=1}^{L_{n-1}} w_{n}^{3,i}(\z) p(\x_n|\x_{n-1}^i,\z)$.
For birth targets, 
this procedure cannot be used 
if the PHD due to birth targets
was computed via an MC approximation.
One can use clustering techniques \cite{VO_SMC_PHD},
or the procedure described in \cite{Ristic-Clark-SMC},
which consists in looking for measurements $\z$ such that
$ \sum_{i=1}^{L_{\gamma_n}} w_{n}^{4,i}(\z)$
is above a given threshold (typically $0.5$);
then an estimator of the state associated to $\z$ is given by
$ \sum_{i=1}^{L_{\gamma_n}} w_{n}^{4,i}(\z) \x_{\gamma_n}^i$.
However, birth targets become persistent targets
at the next time step; 
so their extraction becomes easy at the next iteration 
since an SMC extraction procedure can be avoided.

\begin{remark}
One can also adapt the procedure described above 
\cite{Ristic-Clark-SMC}
to the extraction of persistent target states, 
i.e. looking for measurements
$\z$ such that
$\sum_{i=1}^{L_{n-1}} w_{n}^{3,i}(\z)$ is above a given threshold,
and estimating the associated state by
$\sum_{i=1}^{L_{n-1}} w_{n}^{3,i}(\z) \int \x_n p(\x_n|\x_{n-1}^i,\z){\rm d} \x_n$. 
The advantage of this procedure is
that we just need to compute 
$ \int \x_n p(\x_n|\x_{n-1}^i,\z) {\rm d} \x_n$
for such measurements. 
\end{remark} 

Let us now detail some applications of the CMC-PHD filter.

\subsubsection{Gaussian and linear models 
with Gaussian Mixture (GM) birth intensity: 
an alternative to the GM implementation of the PHD filter}
\label{GM-model}
We first assume that 
$f_{n|n-1}(\x_n|\x_{n-1})=\mathcal{N} (\x_n;{\bf F}_n \x_{n-1}; {\bf Q}_n)$,
$g_n(\z|\x_n)=\mathcal{N} (\z;{\bf H}_n \x_{n}; {\bf R}_n)$, and that
$\gamma_n$ is a GM, i.e. that 
$
\gamma_n(\x_n) \! = \!  \sum_{i=1}^{N_{\gamma_n}} w_{\gamma_n}^i \! \mathcal{N}(\x_n;{\bf m}_{\gamma_n}^i;$ $ {\bf P}_{\gamma_n}^i).
$
For such models a GM implementation has been proposed \cite{VO_GM},
which consists
in propagating a GM approximation of PHD $v_n$ via
\eqref{phd-predic}-\eqref{phd-update}.
The mixture grows exponentially
due to the summation on the set of measurements in
\eqref{phd-update},
so pruning and merging approximations are necessary.
In addition, this implementation
requires that $p_{d,n}$ and $p_{s,n}$
are constant 
(or possibly GM \cite{VO_GM}).
In our algorithm
we do not need to make any assumption
about $p_s(\x_{n-1})$.
For this model
$B_n^2(\z)$ is directly computable,
and the Bayesian CMC procedure 
for estimating the number of targets
and extracting the states is valid 
since $p(\x_n|\x_{n-1},\z)$ and
$p(\z|\x_{n-1})$ are computable (see \eqref{semi-etat}-\eqref{semi-observation} and
\eqref{sampling-distrib}-\eqref{semi-poids}).
Finally, 
in the case where $p_s(\x_{n-1})$ is constant,
we have at our disposal 
three implementations of the PHD filter: 
the GM \cite{VO_GM}, the SMC \cite{VO_SMC_PHD} 
and our Bayesian CMC implementations
which will be compared
in section \ref{simu-PHD} below.

\subsubsection{Gaussian and linear models with ordinary birth intensity}
If $\gamma_n$ is not a GM
the GM implementation cannot be used any longer.
However, our method remains valid
if we compute
$\Theta_{2,n}$, $\Theta_{4,n}$ and $B_n^2(\z)$
via an MC approximation. 
By constrast to the pure SMC technique,
our Bayesian CMC implementation 
enables to keep the GM structure
for persistent targets.

\subsubsection{Non linear models}
In a non-linear model
the GM implementation cannot be used any longer.
The extended (resp. unscented) Kalman 
PHD filter \cite{VO_GM}
approximates the PHD by a GM,
the parameters of which are propagated 
by an EKF (resp. UKF).
By contrast, 
we propose to adapt our Bayesian CMC implementation, 
by approximating $p(\x_n|\x_{n-1},\z)$ 
and $p(\z|\x_{n-1})$ 
at time $n$ by techniques described in \S \ref{computing-practice}.
The main difference is that 
we start from a discrete approximation of the PHD at time $n-1$,
and compute an estimate of the states 
without using clustering techniques of the MC implementation.
This way we get 
an approximation of the PHD 
which does not rely on a numerical approximation
at time $n-1$ and which enables to extract the states easily.
In addition, by contrast to the extended and unscented
implementations of the PHD filter, numerical approximations
are not propagated over time 
since they are only used locally
for the extraction of states.  

\subsection {Simulations}
\label{simu-PHD}

We now compare our Bayesian CMC PHD estimator 
to alternative implementations of the PHD filter.
The MSE criterion used previously 
is not appropriate in the multi-target context: 
since the number of targets evolves,
a performances criterion should take into account
an estimator of the number of targets
and an estimator of their states.
So in this section we will use 
the optimal subpattern assignment (OSPA) distance \cite{SCHUH_OSPA},
which is a classical tool for comparing 
multi-target filtering algorithms.
Let 
$X=\{x_1,...,x_m\}$ and 
$Y=\{y_1,...,y_n\}$ be two finite sets,
which respectively represent
the estimated and true sets of targets.
For $ 1 \leq p < + \infty $ and $c>0$,
let 
$d^{(c)}(x,y)=min(c, ||{x-y}||)$ 
($||.||$ is the euclidean norm)
and 
let $\Pi_n$ be the set of permutations 
on $\left\{1,2,...,n\right\}$.
The OSPA metric is defined by :
\begin{align}
\label{OSPA}
\overline{d}_p^{c}(X,Y) \! \overset{\Delta}{=}  \! \left(\frac{1}{n} \left(\underset {\pi \in \Pi_n}{\min} \sum \limits_{i=1}^{m} d^{(c)}(x_i,y_{ \pi(i)})^p + c^p(n\! - \!m) \right) \right)^{\frac{1}{p}}
\end{align}
if $m \leq n$,
and $\overline{d}_p^{c}(X,Y)\overset{\Delta}{=} \overline{d}_p^{c}(Y,X) $ 
if $m>n$. 
The term $\underset {\pi \in \Pi_n}{\min} \sum \limits_{i=1}^{m} d^{(c)}(x_i,y_{ \pi(i)})^p$
represents the localization error, while the second term represents the cardinality error.


We focus on the linear and Gaussian model
in which the GM-PHD is used as a benchmark solution and 
enables to appreciate the performance of our Bayesian CMC-PHD filter.
So we compare the GM-PHD,
the SMC-PHD and
our Bayesian CMC-PHD filters. 
We track the position and velocity of the targets
so $\x_n=[p_x,\dot{p}_x,p_y,\dot{p}_y]_n^T$.
Let also
$f_{n|n-1}(\x_n|$ $\x_{n-1})=\mathcal{N}(\x_n;{\bf F}_n \x_{n-1}, {\bf Q}_n)$
and $g_n({\bf z}_n|\x_{n})=\mathcal{N}({\bf z}_n;{\bf H}_n \x_{n}, {\bf R}_n)$,
where
${\bf F}_n=\begin{bmatrix}
1 & T & 0 & 0 \\
0 & 1 & 0 & 0 \\
0 & 0 & 1 & T \\
0 & 0 & 0 & 1
\end{bmatrix}$, 
and the other parameters (${\bf H}_n$, ${\bf Q}_n$ and ${\bf R}_n$)
are identical 
to those of \S \ref{simu-jmss}.

We compare the SMC-PHD and our Bayesian CMC filters
in the case where both algorithms
use the transition pdf $f_{n|n-1}(\x_n|\x_{n-1})$
(remember that in our approach,
we need to propagate a discrete approximation of the PHD,
even if it not used for computing 
an estimator of the number of targets). 
We take $T=2s$, $\sigma_v=3m^2/sec^3$ but
$\sigma_x= \sigma_y=0.3m$, which means
that the likelihood $g_n(\z|\x_n)$ is sharp;
since the transition pdf 
does not take into account 
available measurements,
it is difficult to guide particles into promising regions,
so this experimental scenario 
is challenging for the SMC-PHD implementation.
Particles are initialized around the
measurements \cite{Ristic-Clark-SMC}.
In both algorithms we use 
$N_b=20$ particles per newborn target and 
$N=200$ particles per persistent target.
The probability of detection is $p_{d,k}=0.95$
and that of survival $p_{s,k}=0.98$, for all $k$, $1 \leq k \leq 100$,
and we generate $10$ false alarm measurements (in mean).
We consider a scenario with $6$ targets which appear 
either at $k=0$, $k=20$ or $k=50$. 
We also test the GM implementation 
in which 
$T_p=10^{-5}$ for the pruning threshold, 
$T_m=4m$ for the merging threshold
and we keep at most $N_{\rm max}=100$ Gaussians (see \S \ref{GM-model}).

\begin{figure} [htbp]
\center
\epsfig{file=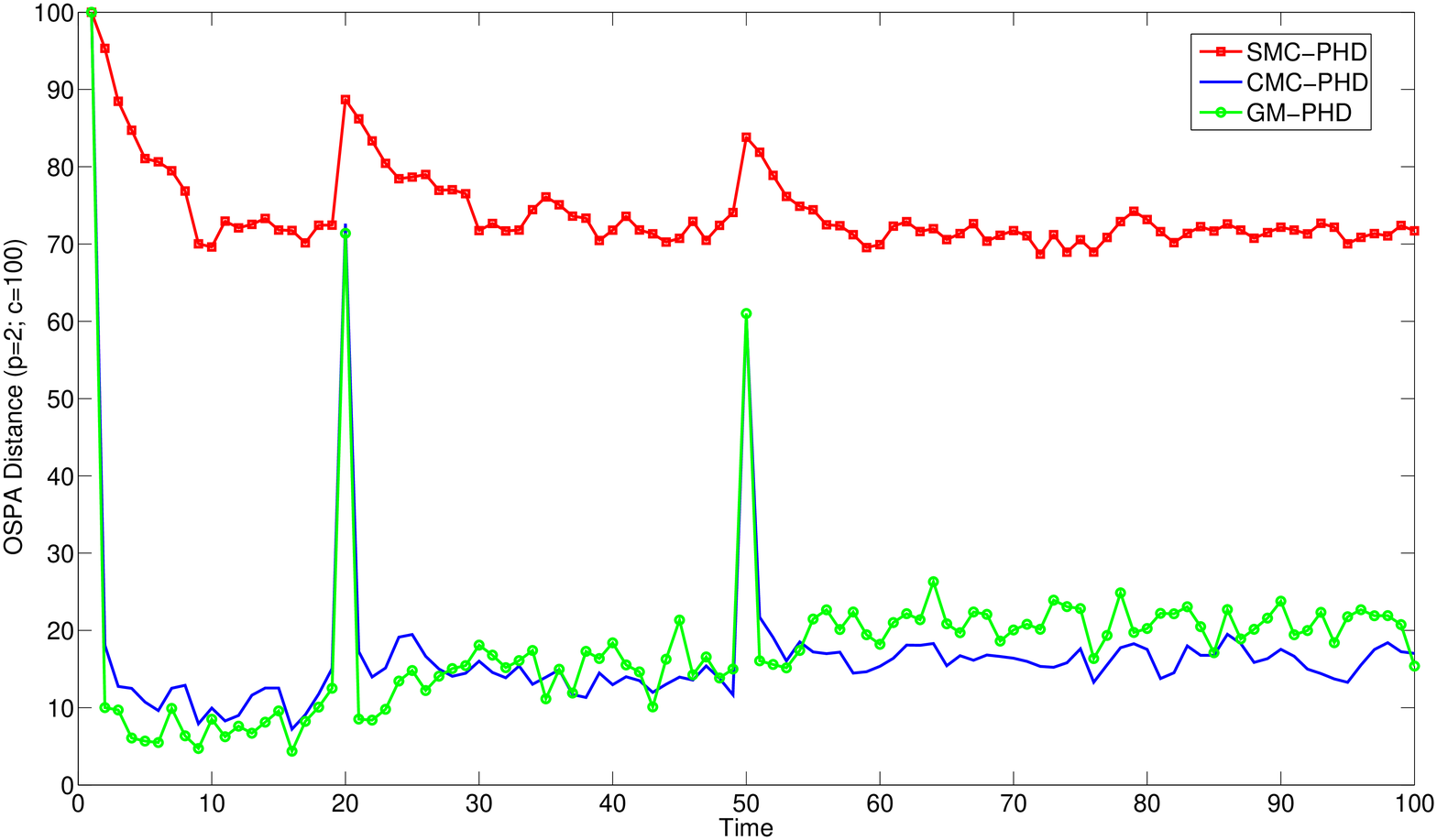,height=0.2\textheight,width=0.85\textwidth}
	\caption{OSPA distance for linear and Gaussian scenario - The GM and CMC implementations widely outperform the classical
	SMC one because of the choice of the likelihood function $g_n(\z | \x)$.}
	\label{OSPA-simu4}
\end{figure}

\begin{figure} [htbp]
\center
\epsfig{file=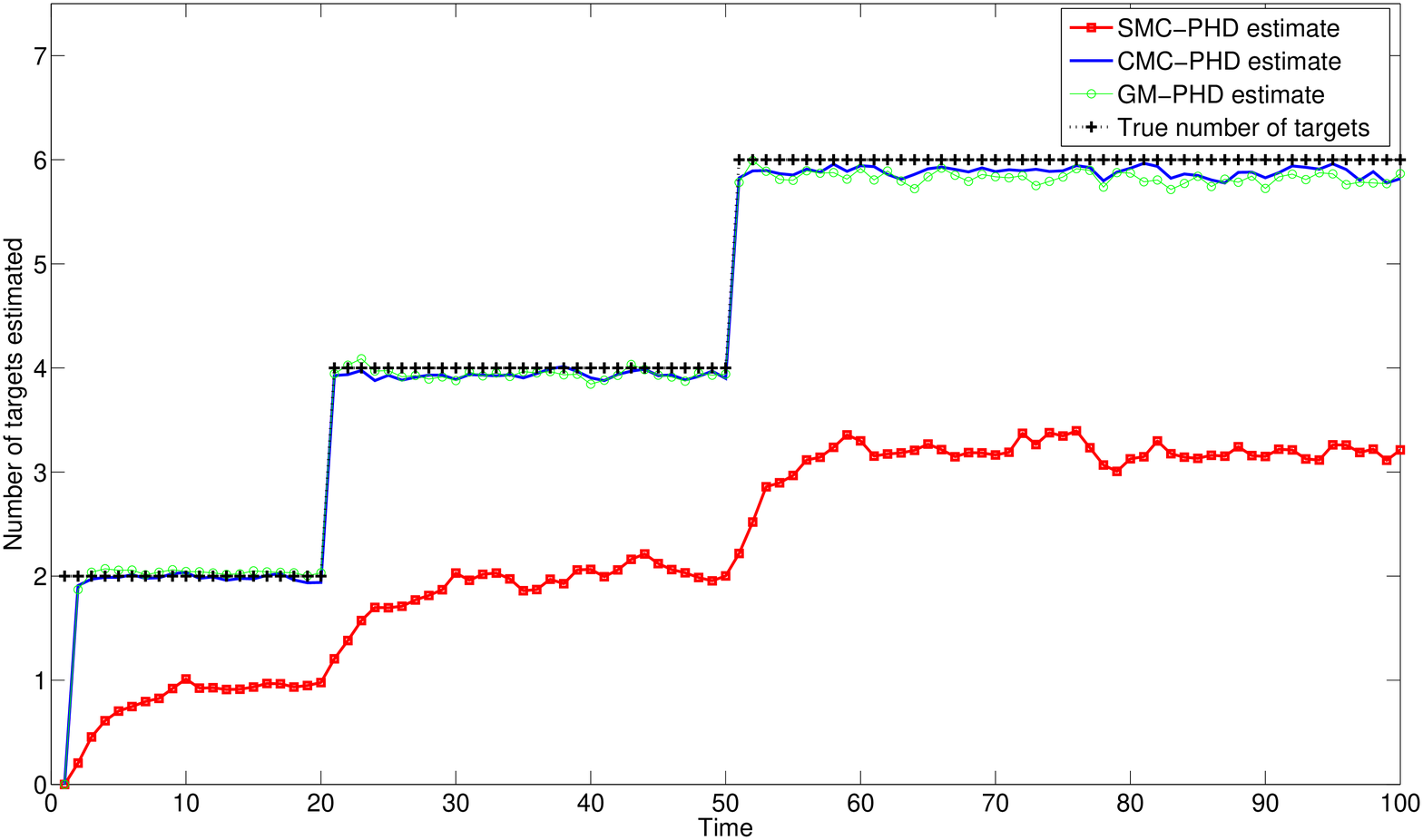,height=0.2\textheight,width=0.85\textwidth}
		\caption{Estimator of the number of targets for linear and Gaussian scenario. }
	\label{nbcibles-simu4}
\end{figure}

\begin{figure} [htbp]
\center
\epsfig{file=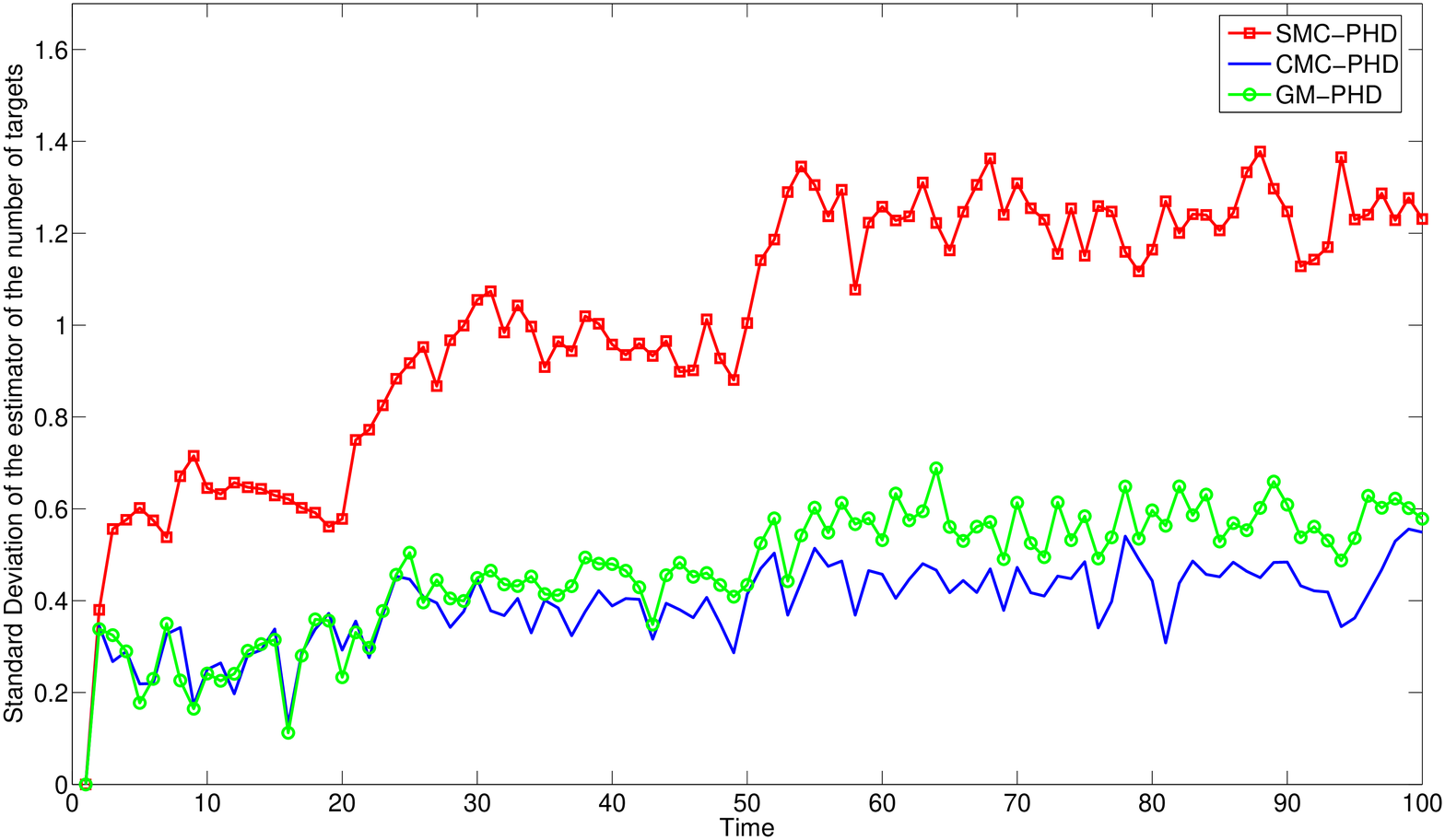,height=0.2\textheight,width=0.85\textwidth}
	\caption{Standard Deviation of the estimator of the number of targets for linear and Gaussian scenario - The CMC estimator
	is slightly more reliable than the GM one when time increases.}
	\label{variance-simu4}
\end{figure}

The OSPA distance and estimated number of targets 
are displayed in Figures 
\ref{OSPA-simu4} and 
\ref{nbcibles-simu4}.
The Bayesian CMC approach outperforms the SMC one 
and copes with the issue of guiding particles in promising regions.
Even if we use the transition density
for getting a discrete approximation of 
$v_{n-1}$, 
the Bayesian CMC approach provides
a correct estimate of the number of targets,
by contrast to the SMC one in which
the new set $\{\x_n^i,w_n^i\}_{i=1}^{L_n}$
is used to deduce 
a discrete approximation of $v_{n}$,
then an estimate of the number of targets. 
The Bayesian CMC PHD estimator
also outperforms the GM one
in terms of OSPA distance.
Finally the number of targets is
well estimated 
both by the GM and Bayesian CMC implementations, 
but the Bayesian CMC estimator
is more accurate, 
see Figure \ref{variance-simu4}.

\section{Conclusion}
In this paper 
we adapted CMC to single- and multi-object Bayesian filtering.
In this framework, 
the recursive nature of SMC algorithms 
provides a conditioning variable 
at each time instant,
but i.i.d. samples from this conditioning variable are unavailable.
Our variance reduction method
can be seen as a temporal, rather than spatial,
RB-PF procedure;
a Bayesian CMC estimator is ensured to
outperform the associated crude MC one
whatever the number of particles.
We next showed that a CMC estimator 
can indeed be computed, or approximated,
in a variety of Markovian stochastic models,
including semi-linear HMC or JMSS,
either at the same cost
or at a reasonable extra computational cost.
Finally we adapted Bayesian CMC to multi-target filtering,
and showed that our CMC PHD filter 
has interesting practical features 
as compared to alternate (SMC or GM)
implementations of the PHD filter.
Our analysis was validated via simulations.

\section{Aknowledgements}
The authors would like to thank the French MOD DGA/MRIS for financial
support of the Ph.D. of Y. Petetin.

\bibliographystyle{elsarticle-num}
\bibliography{yohan-p}

\begin{thebibliography}{10}
\expandafter\ifx\csname url\endcsname\relax
  \def\url#1{\texttt{#1}}\fi
\expandafter\ifx\csname urlprefix\endcsname\relax\def\urlprefix{URL }\fi
\expandafter\ifx\csname href\endcsname\relax
  \def\href#1#2{#2} \def\path#1{#1}\fi

\bibitem{livredoucet}
A.~Doucet, N.~de~Freitas, N.~Gordon, Sequential Monte Carlo Methods in
  Practice, Statistics for Engineering and Information Science, Springer
  Verlag, New York, 2001.

\bibitem{arulampalamshort}
M.~S. Arulampalam, S.~Maskell, N.~Gordon, T.~Clapp, A tutorial on particle
  filters for online nonlinear / non-{G}aussian {B}ayesian tracking, IEEE Tr.
  Signal Processing 50~(2) (2002) 174--188.

\bibitem{MAHLER_ARTICLE2003}
R.~Mahler, Multitarget {B}ayes filtering via first-order multitarget moments,
  IEEE Transactions on Aerospace and Electronic Systems 39~(4) (2003)
  1152--1178.

\bibitem{asmussen-glynn}
S.~Asmussen, P.~W. Glynn, \textit{Stochastic Simulation: Algorithms and
  Analysis}, Springer-Verlag, New York, NY, 2007.

\bibitem{Chen-Mixture-Kalman}
R.~Chen, J.~S. Liu, Mixture {K}alman filters, J. R. Statist. Soc. B 62 (2000)
  493--508.

\bibitem{doucet-sequentialMC}
A.~Doucet, S.~J. Godsill, C.~Andrieu, On sequential {M}onte {C}arlo sampling
  methods for {B}ayesian filtering, Statistics and Computing 10 (2000)
  197--208.

\bibitem{doucet-jump-Mkv}
A.~Doucet, N.~J. Gordon, V.~Krishnamurthy, Particle filters for state
  estimation of jump {M}arkov linear systems, IEEE Transactions on Signal
  Processing 49~(3) (2001) 613--24.

\bibitem{Shon-MPF}
T.~Sch\"on, F.~Gustafsson, P.-J. Nordlund, Marginalized particle filters for
  mixed linear nonlinear state-space models, IEEE Trans. on Signal Processing
  53 (2005) 2279--2289.

\bibitem{Gewecke}
J.~Gewecke, Bayesian inference in econometric models using {M}onte {C}arlo
  integration, Econometrica 57~(6) (1989) 1317--1339.

\bibitem{chopin}
N.~Chopin, Central limit theorem for sequential {M}onte {C}arlo methods and its
  application to {B}ayesian inference, The Annals of Statistics 32~(6) (2004)
  2385--2411.

\bibitem{lindstein-rao}
F.~Lindsten, T.~Sch\"on, J.~Olsson, An explicit variance reduction expression
  for the {R}ao-{B}lackwellized particle filter, in: 18th World Congress of the
  Int. Federation of Automatic Control (IFAC), 2011.

\bibitem{rubin1988}
D.~B. Rubin, Using the {SIR} algorithm to simulate posterior distributions, in:
  M.~H. Bernardo, K.~M. Degroot, D.~V. Lindley, A.~F.~M. Smith (Eds.), Bayesian
  Statistics III, Oxford University Press, Oxford, 1988.

\bibitem{gelfand-smith}
A.~E. Gelfand, A.~F.~M. Smith, Sampling based approaches to calculating
  marginal densities, Journal of the American Statistical Association 85~(410)
  (1990) 398--409.

\bibitem{smith-gelfand}
A.~F.~M. Smith, A.~E. Gelfand, Bayesian statistics without tears : a
  sampling-resampling perspective., The American Statistician 46~(2) (1992)
  84--87.

\bibitem{zaritskii1975}
V.~Zaritskii, V.~Svetnik, L.~Shimelevich, Monte {c}arlo techniques in problems
  of optimal data processing, Automation and remote control (1975) 95--103.

\bibitem{kong1994}
A.~Kong, J.~S. Liu, W.~H. Wong, Sequential imputations and {B}ayesian missing
  data problems, Journal of the American Statistical Association 89~(425)
  (1994) 278--88.

\bibitem{liu-chen1995}
J.~S. Liu, R.~Chen, Blind deconvolution via sequential imputation, Journal of
  the American Statistical Association 90~(430) (1995) 567--76.

\bibitem{Saha_EMM}
S.~Saha, P.~K. Manda, Y.~Boers, H.~Driessen, A.~Bagchi, Gaussian proposal
  density using moment matching in {SMC} methods., Statistics and Computing
  19-2 (2009) 203--208.

\bibitem{julier-procieee}
S.~Julier, J.~Uhlmann, Unscented filtering and nonlinear estimation, in:
  Proceedings of the IEEE, Vol.~92, 2004, pp. 401--422.

\bibitem{auxiliary}
M.~K. Pitt, N.~Shephard, Filtering via simulation : Auxiliary particle filters,
  Journal of the American Statistical Association 94~(446) (1999) 590--99.

\bibitem{fearnhead}
P.~Fearnhead, Computational methods for complex stochastic systems: A review of
  some alternatives to {MCMC}, Statistics and Computing 18~(2) (2008) 151--71.

\bibitem{doucet-APF}
A.~M. Johansen, A.~Doucet, A note on the auxiliary particle filter, Statistics
  and Probability Letters 78~(12) (2008) 1498--1504.

\bibitem{Cappeetal}
O.~Capp{\'e}, {\'E}.~Moulines, T.~Ryd{\'e}n, Inference in Hidden {M}arkov
  Models, Springer-Verlag, 2005.

\bibitem{gordon-salmond-smith}
N.~J. Gordon, D.~J. Salmond, A.~F.~M. Smith, Novel approach to nonlinear/
  non-{G}aussian {B}ayesian state estimation, IEE Proceedings-F 140~(2) (1993)
  107--113.

\bibitem{tugnait}
J.~K. Tugnait, Adaptive estimation and identification for discrete systems with
  {M}arkov jump parameters, IEEE Transactions on Automatic Control 27~(5)
  (1982) 1054--65.

\bibitem{shalom-li-kirubarajan}
Y.~Bar-Shalom, X.~R. Li, T.~Kirubarajan, Estimation with {A}pplications to
  {T}racking and {N}avigation, John Wiley and sons, New-York, 2001.

\bibitem{musso-oudjane-legland}
C.~Musso, N.~oudjane, F.~LeGland, Improving regularised particle filters, in:
  A.~Doucet, N.~de~Freitas, N.~Gordon (Eds.), Sequential Monte Carlo Methods in
  Practice, Statistics for Engineering and Information Science, Springer
  Verlag, New York, 2001.

\bibitem{ginnity}
S.~McGinnity, G.~W. Irwin, Multiple model bootstrap filter for maneuvering
  target tracking, \textit{IEEE Transactions on Aerospace and Electronic
  Systems } 36~(3) (2000) 1006--1012.

\bibitem{Andrieux-JM}
C.~Andrieu, M.~Davy, A.~Doucet, Efficient {P}article {F}iltering {F}or {J}ump
  {M}arkov {S}ystems, IEEE trans. on Signal Processing 51 (2002) 1762--1770.

\bibitem{lecuyer}
P.~L'Ecuyer, Efficiency improvement and variance reduction, in: Winter
  Simulation Conference 1994, 1994, pp. 122--132.

\bibitem{SHALOM_JPDAshort}
Y.~Bar-Shalom, Tracking and data association, Academic Press Professional,
  Inc., San Diego, CA, 1987.

\bibitem{BLACKMAN}
S.~Blackman, R.~Popoli, \textit{Design and {A}nalysis of {M}odern {T}racking
  {S}ystems}, Artech House, 2009.

\bibitem{MAHLER_LIVRE2007}
R.~Mahler, Statistical Multisource Multitarget Information Fusion, Artech
  House, 2007.

\bibitem{VO_SMC_PHD}
B.-N. Vo, S.~Singh, A.~Doucet, Sequential {M}onte {C}arlo methods for
  multi-target filtering with random finite sets, IEEE Transactions on
  Aerospace and Electronic Systems 41.

\bibitem{Ristic-Clark-SMC}
B.~Ristic, D.~Clark, B.~Vo, Improved {SMC} implementation of the {PHD} filter,
  in: Proceedings of the 13th International Conference on Information Fusion,
  2010.

\bibitem{VO_GM}
B.-N. Vo, W.~Ma, The {G}aussian mixture probability hypothesis density filter,
  IEEE Transactions on Signal Processing 54 (2006) 4091--4104.

\bibitem{SCHUH_OSPA}
D.~Schuhmacher, B.~T. Vo, B.~N. Vo, A consistent metric for performance
  evaluation of multi-object filters, IEEE Transactions on Signal Processing
  56~(8) (2008) 3447--3457.

\end{thebibliography}

\end{document}